\def\pf{\phantom{$-$}}
\def\av#1{\langle#1\rangle}
\documentclass[preprint,aps,preprintnumbers,amsmath,amssymb]{revtex4-1}
\usepackage[]{graphicx}
\usepackage{latexsym}
\usepackage{natbib}
\usepackage{amsmath}
\usepackage[caption=false]{subfig}

\begin{document}

\title{Avoiding catastrophic failure in correlated
network of networks}

\author{Saulo D. S. Reis$^{1,2}$, Yanqing Hu$^{1}$, Andr\'es Babino$^{3}$,
Jos\'e~S.~Andrade~Jr.$^{2}$, Santiago Canals$^{4}$, Mariano
  Sigman$^{3,5}$, Hern\'an A. Makse$^{1,2,3}$}

\affiliation{ $^1$ Levich Institute and Physics Department, City College of New
  York, New York, New York 10031, USA \\ $^2$ Departamento de
F\'isica, Universidade Federal do Cear\'a, 60451-970 Fortaleza,
Cear\'a, Brazil\\ $^3$ Departamento de F\'isica, FCEN-UBA, Ciudad
Universitaria, (1428) Buenos Aires, Argentina \\ $^4$ Instituto de
Neurociencias, CSIC-UMH, Campus de San Juan, Avenida Ram\'on y Cajal,
03550 San Juan de Alicante, Spain\\ $^5$ Universidad Torcuato Di
Tella, S\'aenz Valiente 1010, C1428BIJ Buenos Aires, Argentina }


\begin{abstract}
{\bf Networks in nature do not act in isolation but instead exchange
information, and depend on each other to function
properly \cite{little,rosato,Buldyrev2010}.  An incipient theory of
Networks of Networks have shown that connected random networks may
very easily result in abrupt failures
\cite{Buldyrev2010,raissa,raissa2,gao}. This theoretical finding
bares an intrinsic paradox~\cite{ref2,ref3}: If natural systems organize in
interconnected networks, how can they be so stable?  Here we provide a
solution to this conundrum, showing that the stability of a system of
networks relies on the relation between the internal structure of a
network and its pattern of connections to other networks.
Specifically, we demonstrate that if network inter-connections are
provided by hubs of the network and if there is a moderate degree of
convergence of inter-network connection the systems of network are
stable and robust to failure. We test this theoretical prediction in
two independent experiments of functional brain networks (in task- and
resting states) which show that brain networks are connected with a
topology that maximizes stability according to the theory.  }
\end{abstract}

\maketitle


Over the last decade the science of complex networks has flourished,
describing the organization of a myriad of natural systems including
societies, Internet, the brain and cell organization, as a web of
interacting nodes~\cite{Book}.  This research program demonstrated
that many critical properties of a system organization, growth and
robustness, depend on how nodes are interconnected and are relatively
independent of the specific identity of each node.

More recently, this argument has been pushed further. Nodes organize
into networks, but these emergent systems do not occur in isolation
from other networks.  Instead, more often networks exchange
information, and depend on each other to function
properly \cite{Buldyrev2010,raissa,raissa2,gao}.  A paradigmatic
example is the power and communication
networks \cite{rosato,little,Buldyrev2010,gao}: communication network
nodes rely for power supply on the power stations and, reciprocally,
the power stations function properly exchanging information through
the communication network.  Understanding how stability and
information flow are affected by these inter-dependencies constitutes
a major challenge to understand the resilience of natural systems.

The theory of networks of networks has been built relying mainly on
unstructured patterns of between-networks connectivity, namely with
one-to-one random interconnections between dependent
nodes \cite{Buldyrev2010,gao}. When two stable networks are fully
interconnected with one-to-one random connections where every single
node in a network depends on a node in the other network chosen at
random, the interaction results in abrupt
failures \cite{Buldyrev2010,gao}: small perturbations in one network
are amplified on an interconnected network, which causes further
damage to the originally perturbed network. This process leads to
cascading failures which are argued to underlay catastrophic outcomes
in man-made infrastructures such as blackouts in power
grids~\cite{Buldyrev2010,rosato}. However, this theoretical finding
bares an intrinsic paradox \cite{ref2,ref3}: If living systems--- such
as the brain \cite{Dosenbach2007} and cellular
networks \cite{vidal}--- organize in interconnected networks, how can
they be so stable?

Our conjecture is that the solution to this conundrum relies on the
relation between the internal structure of the set of networks and its
pattern of connections to other networks.  Random networks are very
efficient mathematical constructs to develop theory but the majority
of networks observed in nature are
correlated \cite{vespignani,gallos08}. Correlations, in turn, provide
structure to the network.  Indeed, the importance of degree
correlations on the dynamical and structural properties of
interconnected networks has been recently addressed in
Ref. \cite{ref1}.

Most natural networks form hubs which make certain nodes of greater
relevance. This structure adds a degree of freedom to the system of
networks by setting whether hubs of the network should be the nodes
broadcasting information to other networks, or, conversely, whether
across networks communication should be governed by nodes with less
pertinence within the network.

We develop a full theory for systems of structured networks which
identifies a structural communication protocol which assures that the
system of networks is stable (less likely to break into catastrophic
failures) and optimized for fast communication across the entire
system. The theory establishes concrete predictions of a regime of
correlated connectivity between networks composing the system.

We test these predictions with two different systems of brain
connectivity based on functional magnetic resonance imaging (fMRI)
data. The brain organizes in a series of interacting
networks~\cite{Dosenbach2007,Gallos2012} presenting a paradigmatic
case study for a theory of connected correlated networks. We show that
for two independent experiments of functional networks in task and
resting state in humans, the systems of brain networks organize
optimally as predicted by the theory.

Our results hence provide a plausible explanation to {\it (i)} the
conundrum of why systems of networks were theoretically expected to
show frequent catastrophic failure but this was not observed in
nature, {\it (ii)} provide a specific theoretical prediction on how
structured networks should be interconnected to be stable, and {\it
(iii)} demonstrate in two examples of functional brain connectivity
that the structure of across network connections lies in the range
where the theory predicts stability for different functional
architectures.

We present a theory based on a recursive set of equations to study the
cascading failure and percolation process for two correlated
interconnected networks.  The theory is a generalization of the
analytical approach for single networks of Moore and
Newman \cite{Moore2000} to study cascading behavior in interconnected
correlated networks (analytic details in SI
Section~\ref{generating}). Here we refer to the most important aspects
of the theory and the corresponding set of predictions.  The
theory can be extended to n-interconnected networks following
Ref. \cite{gao2}.

We consider two interconnected networks, each one has a power-law
degree distribution characterized by exponent $\gamma$, $P(k_{\rm
in})\sim k_{\rm in}^{-\gamma}$, valid up to a cut-off $k_{\rm max}$
imposed by their finite size.  Here $k_{\rm in}$ is the number of
links of a node towards nodes in the same network. This implies that a
few nodes will be vastly connected within the network (hubs) while the
majority of nodes will be weakly connected to other nodes in the
network. 

\begin{figure*}
  \begin{center}
   \includegraphics[width=.6\columnwidth]{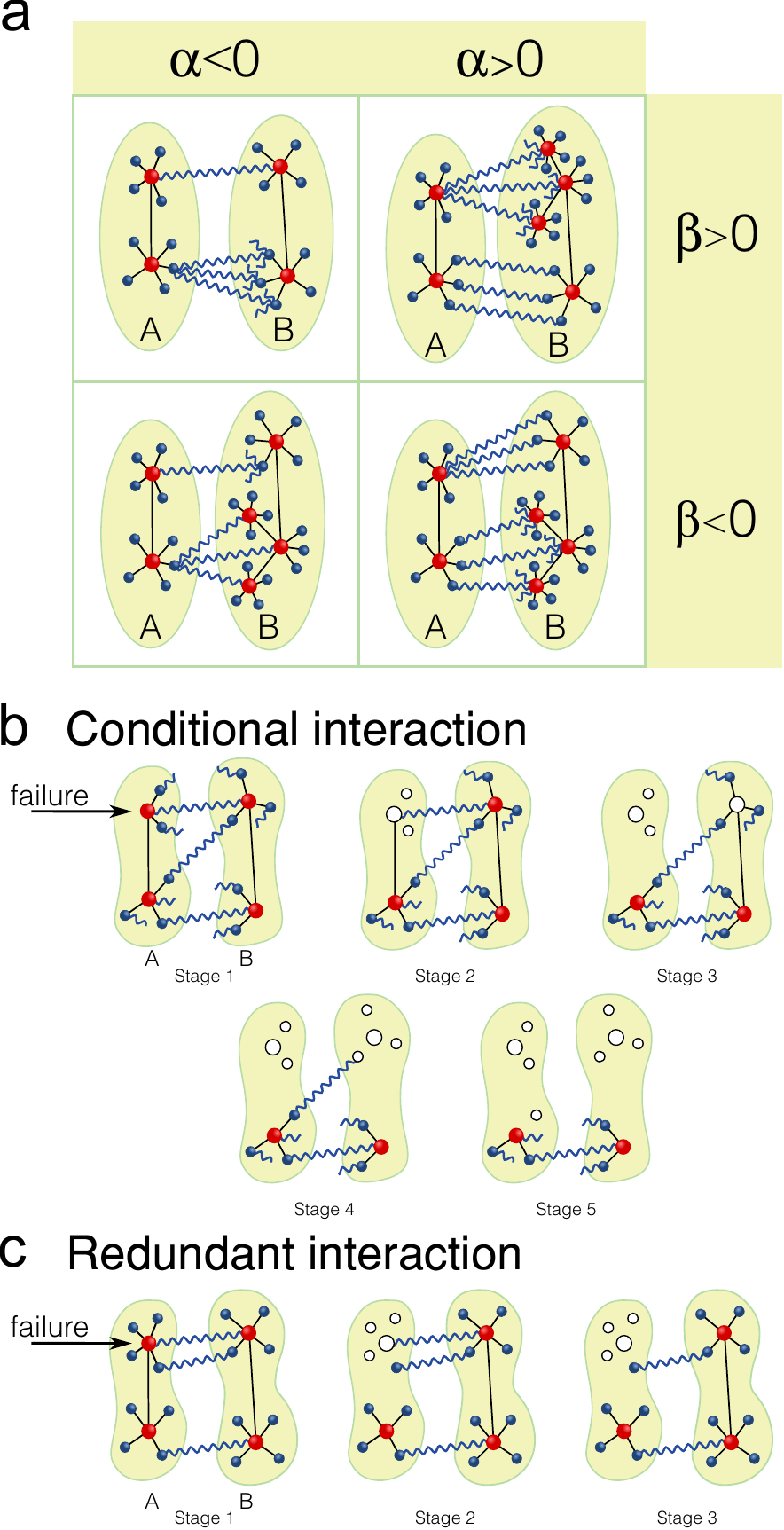}
  \end{center}
 \caption{}
 \label{fig1.fig}
 \end{figure*}

The structure between interconnected networks can be characterized by
two parameters: $\alpha$ and $\beta$ (Fig.~\ref{fig1.fig}a). The
parameter $\alpha$, defined as
\begin{equation}
k_{\rm out} \sim k_{\rm in}^\alpha,
\label{alpha.eq}
\end{equation}
where $k_{\rm out}$ is the degree of a node towards nodes in the other
network, determines the likelihood that hubs of each network are also
the principal nodes connecting both networks. For $\alpha>0$ the nodes
in network $A$ and $B$ which connect both networks will typically be
hubs in $A$ and $B$ respectively (Fig. ~\ref{fig1.fig}a, right
panels). Instead, for $\alpha<0$ the two networks will be connected
preferentially by nodes of low degree within each network
(Fig.~\ref{fig1.fig}a, left panels). 

The parameter $\beta$ defines the indegree-indegree internetwork
correlations as \cite{vespignani,gallos08}:
\begin{equation}
k^{\rm nn}_{\rm in} \sim k_{\rm in}^\beta,
\label{beta.eq}
\end{equation}
where $k^{\rm nn}_{\rm in}$ is the average in-degree of the
nearest-neighbors of a node in the other network.  It determines the
convergence of connections between networks, i.e. the likelihood that
a link connecting networks $A$ and $B$ coincides in the same type of
node. Intuitively, Eqs. (\ref{alpha.eq})-(\ref{beta.eq}) can be seen
as a compromise between redundancy and reach of connections between
both networks. For $\beta > 0$ connections between networks are
convergent (assortative, Fig.~\ref{fig1.fig}a, top panels), while for
$\beta < 0$ they are divergent (dissasortative, Fig. \ref{fig1.fig}a,
bottom panels).  Uncorrelated networks have $\alpha=0$ and $\beta=0$.

We analyze how the system of two correlated networks breaks down after
random failure (random attack) of a fraction $1-p$ nodes for different
patterns of between-networks connectivity characterized by
$(\alpha,\beta)$. We adopt the conventional percolation criterion of
stability and connectivity measuring how the largest connected
component breaks-down following the attack~\cite{Buldyrev2010}.  In
classic percolation of single networks, two nodes of a network are
randomly linked with probability $p$~\cite{bollobas}. For low $p$, the
network is fragmented into subextensive components. Percolation theory
of random networks demonstrates that as $p$ increases, there is a
critical phase transition in which a single extensive cluster or giant
component spans the system (the critical $p$ is referred to as $p_c$).

A robust notion of stability in a system of networks can be obtained
by identifying $p_c$ at which a cohesive mutually connected network
breaks down into disjoint sub-components under different forms of
attack. Network topologies with low $p_c$ are robust, since this
indicates that the majority of nodes ought to be removed to break it
down. On the contrary, high values of $p_c$ are indicative of a
fragile network which breaks down by only removing a few nodes.

Here we analyze two qualitatively different manners in which the
networks interact and propagate failure.  In one mode ({\it
conditional interaction}, Fig. \ref{fig1.fig}b) a node in network $B$
cannot function (and hence is removed) if it looses all connectivity
with network $A$ after the attack~\cite{Buldyrev2010}. In the second
condition ({\it redundant interaction}, Fig. \ref{fig1.fig}c) a node
in network $B$ may survive even if it is completely decoupled from
network $A$, if it remains attached to the largest component of
network $B$~\cite{raissa}. To understand why these two responses to
failure are pertinent in real networks it helps to exemplify the
interaction between power and data networks.  If electricity can only
flow through the cables of the power network, a node in the data
network unplugged from the power system shuts off and stops
functioning. This situation corresponds to two networks coupled in a
conditional manner; a case treated in Ref. ~\cite{Buldyrev2010}
considering one-to-one random connections between networks. Consider
instead the case of a printer or any peripheral which can be plugged
to the main electricity network but can also receive power through a
USB cable by the computer. A node may still function even if it is
disconnected from the other network, if it remains connected to its
local network. This corresponds to the redundant interaction as
treated by Ref. ~\cite{raissa} in the unstructured case.

\begin{figure*}
 \begin{center}
   \includegraphics[width=.85\columnwidth]{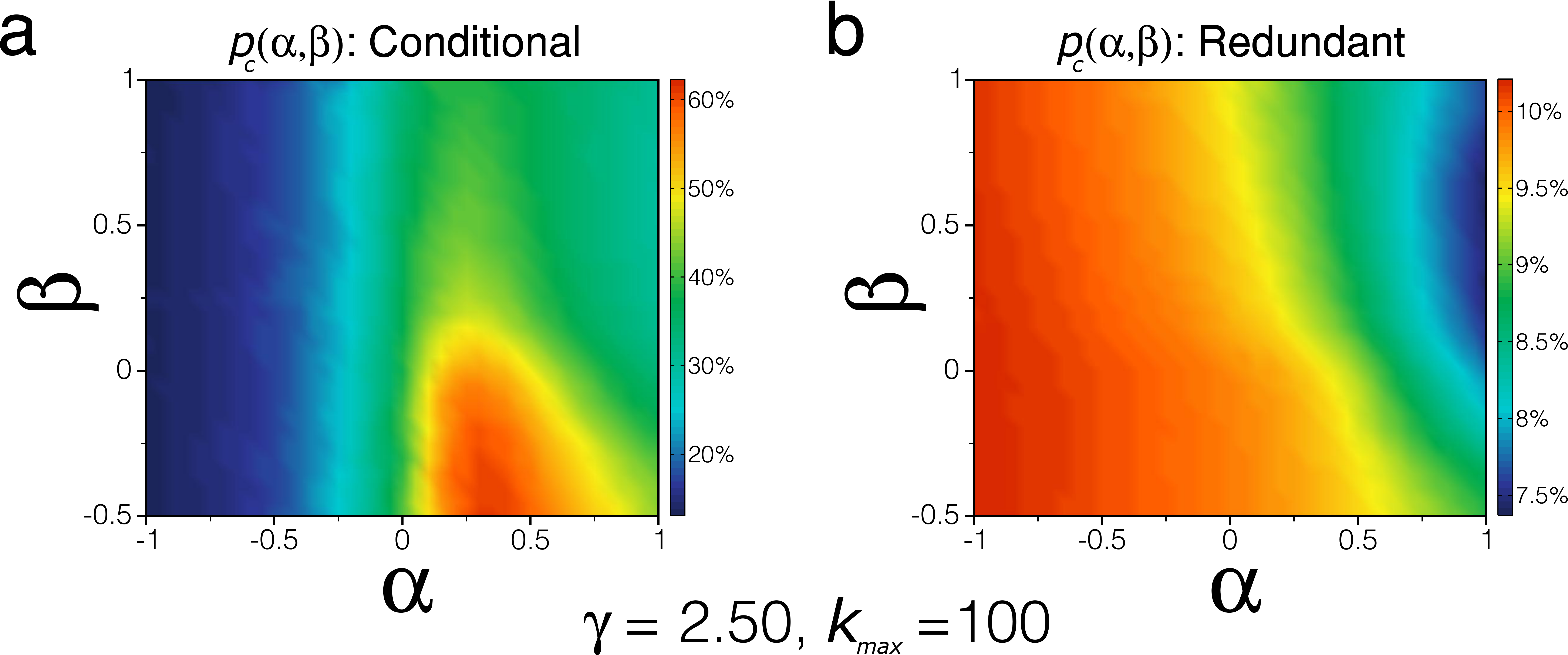}
  \end{center}
 \caption{}
 \label{fig2.fig}
 \end{figure*}

We first investigate the stability of two interacting scale-free
networks for a value of $\gamma$ set arbitrarily to $2.5$ and $k_{\rm
max}=100$ in a regime where each isolated network is stable and robust
to attack ~\cite{Cohen2002}.
The attack starts with the removal of a fraction of $1-p$ nodes chosen
at random from both networks. This attack produces extra failures of,
for instance, nodes in $B$, if
{\it (i) conditional interaction:} they disconnect from the giant
component of network $A$ {\it or} disconnect from the giant component
of $B$, or {\it (ii) redundant interaction:} they disconnect from the
giant component of network $A$ {\it and} the giant component of
network $B$. In conditional mode, this process may lead to new
failures in network $A$ producing a cascade if they loose connectivity
in $B$. Other nodes in $A$ may also fail as they get disconnected from
the giant component in $A$, and the cascading process iterates until
converging to a final configuration. By definition, only the
conditional mode may produce cascading effects but not the redundant
mode. The theoretical analysis of this process leads to a set of
recursive equations (SI Section ~\ref{generating}) that provides a
stability phase diagram for the critical percolation threshold
$p_c(\alpha,\beta)$ under attack in redundant and conditional failures
for a given $(\gamma,k_{\rm max})$ as seen in Fig.~\ref{fig2.fig}.

Figure \ref{fig2.fig} reveals that the relation between a network
internal structure and the pattern of connection between networks
critically determines whether attacks lead to catastrophic cascading
failures (high $p_c$) or not (low $p_c$). For conditional
interactions, the system of networks is stable when $\alpha < 0$
(indicated by low $p_c(\alpha,\beta)$, left-blue region in
Fig. \ref{fig2.fig}a) or for $\alpha \gtrsim 0.5$ and $\beta > 0$
(light blue top-right quadrant), and becomes particularly unstable for
intermediate values of $0<\alpha<0.5$ and $\beta<0$. This result shows
that the system of networks is stable when the hubs are protected
$\alpha<0$ by being isolated from network-network connectivity or
when, on the contrary, the bulk of connectivity within and across
networks is sustained exclusively by a very small set of hubs (large
$\alpha, \beta$). Intermediate configurations where hubs interconnect
with low-degree nodes, are highly unstable since hubs can be easily
attacked via conditional interactions, and lead to catastrophic
cascading after attack. Similar unstable configurations appear in the
one-to-one random interconnectivity \cite{Buldyrev2010}.

When two networks interact in a redundant manner, the system of
networks is less vulnerable to attacks (Fig.~\ref{fig2.fig}b).  This
expected result is manifested by the fact that even for small values
of $p\sim0.1$, the system of networks remains largely connected for
any $(\alpha,\beta)$. The non-intuitive observation is that the
relation between a network internal structure and the pattern of
connection between networks which optimizes stability differs from the
conditional interaction (Fig.~\ref{fig2.fig}a). In fact, $\alpha < 0$
leads to the less stable configurations (larger value of $p_c$ in
Fig. \ref{fig2.fig}b, red region), and the only region which maximizes
stability corresponds to high values of $\alpha$ and $\beta>0$ (blue
region in Fig. \ref{fig2.fig}b), i.e. an interaction where connection
between networks is highly redundant and carried only by a few hubs of
each network. Thus, the parameters that maximize stability for both
interactions lie in the region $\alpha\approx 1$ and $\beta>0$.

Systems of brain networks present an ideal candidate to examine this
theory for the following reasons: {\it (i)} Local-brain networks
organize according to a power-law degree
distribution \cite{eguiluz,achard}, and {\it (ii)} some aspects of
local function are independent of long-range global interactions with
other networks (as in the redundant interaction) like the processing
of distinct sensory features, while other aspects of local
connectivity can be shut-down when connectivity to other networks is
shut-down (as in conditional interaction) like integrative perceptual
processing \cite{gilbert_sigman_top_down}.  Hence, the theory predicts
that to assure stability for both modes of dependencies, brain
networks ought to be connected with positive and high values of
$\alpha$ and positive values of $\beta$.

\begin{figure*}
  \begin{center}
   \includegraphics[width=.8\columnwidth]{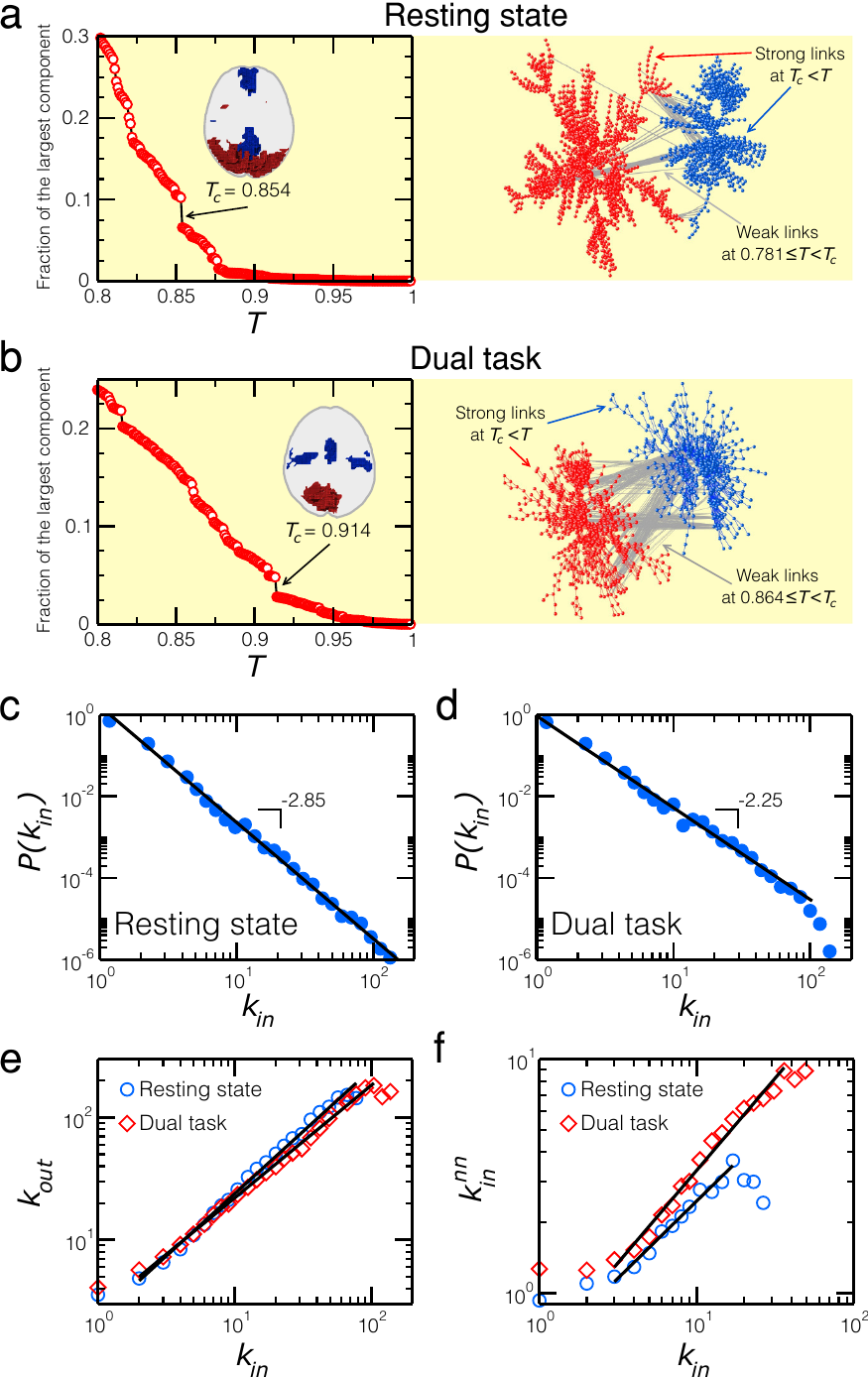}
  \end{center}
 \caption{}
 \label{fig3.fig}
 \end{figure*}

In the next section we examine this hypothesis for two independent
functional magnetic resonance imaging (fMRI) experiments:
human-resting state data obtained from NYU public
repository \cite{Shehzad2009} and human dual-task
data \cite{Sigman2008} previously used to investigate brain network
topology \cite{Gallos2012,lucilla,frontiers} (see Methods Section and
SI Section \ref{experiments} for details).  We first identify
functional networks (in resting state, Fig.~\ref{fig3.fig}a and dual
task, Fig. \ref{fig3.fig}b) made of nodes connected by strong links,
ie, by highly correlated fMRI signals \cite{Gallos2012}. These
networks are interconnected by weak links (low-correlation in the fMRI
signal) following the clustering methods of
Ref. \cite{Gallos2012}. The indegree distribution of the system of
networks follows a bounded power-law (Fig. \ref{fig3.fig}c-d and
Table \ref{exp.table}) and the exponents $\alpha$ and $\beta$ show
high positive values for both experiments (Fig.~\ref{fig3.fig}e-f and
Table \ref{exp.table}).

\begin{figure*}
  \begin{center}
   \includegraphics[width=.85\columnwidth]{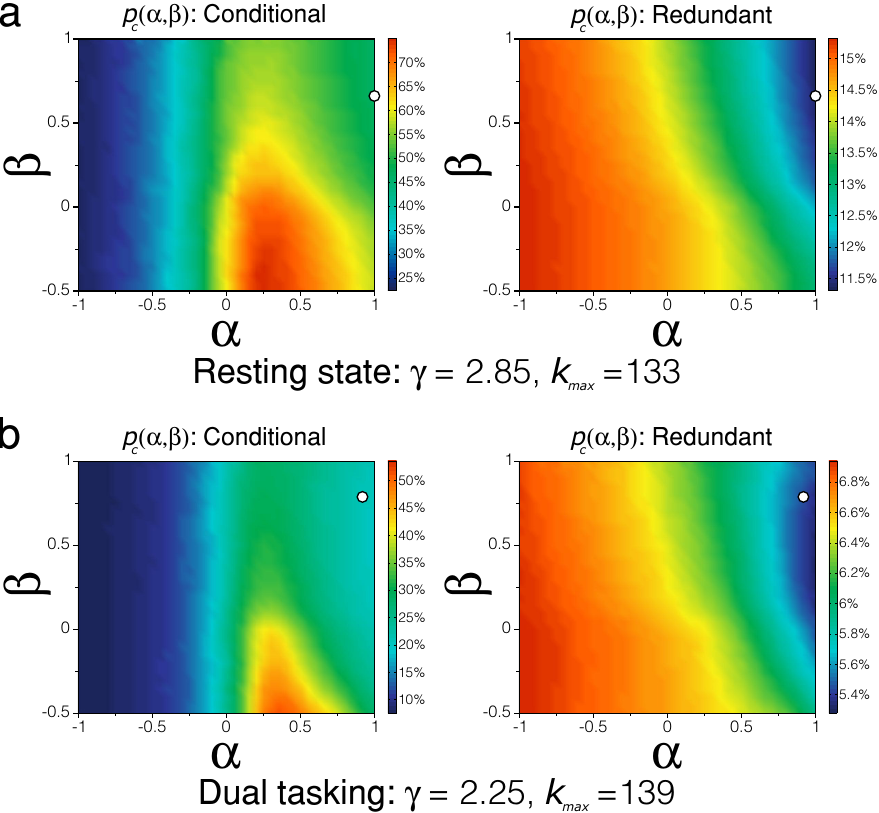}
  \end{center}
 \caption{}
 \label{fig4.fig}
 \end{figure*}

To examine whether these values are optimal for the specific
$(\gamma,k_{\rm max})$-parameters of these networks, we projected for
each experiment, the measured values of $\alpha$ and $\beta$ to the
theoretically constructed stability phase diagram quantified by
$p_c(\alpha,\beta)$ in conditional and redundant mode
(Fig.~\ref{fig4.fig}).
Remarkably, the experimental values of $\alpha$ and $\beta$ (white
circles) lie within the relatively narrow region of parameter space that
minimizes failure for conditional and redundant interaction. Overall
these results demonstrate that brain networks tested under distinct
mental states share the topological features that confer stability to
the system.

Our result hence provides a theoretical revision to the current view
that systems of networks are highly unstable. We show that for
structured networks, if the inter-connections are provided by hubs of
the network ($\alpha>0.5$) and for moderate degrees of convergence of
inter-network connection ($\beta>0$) the systems of network are
stable. This stability holds in the conditional
interaction~\cite{Buldyrev2010} and in a more robust topology of
redundant interaction~\cite{raissa}. The redundant condition is
equivalent to stating that the system of networks merges in a single
network (in-going and out-going links are treated as the same). Hence
the condition of optimality for this topology equates to saying that
the size of the giant component formed by the connection of both
networks is optimized.  As a consequence, the maximization of
robustness for both conditions is equivalent to maximize {\it (i)}
robustness in the more conventional conditional interaction, where
links of one network are strictly necessary for proper function of the
other network, and {\it (ii)} a notion of information flow and storage
using classic percolation theory definition of the size of the maximal
mutual component across both networks. In other words, these
parameters form a set of interacting nodes which are maximally large
in size and robust to failure.

The most natural metaphor for man-made system of networks is for
electricity (wires) and the Internet or voice connectivity (data). A
more direct analogue to this case in a living system such as the brain
would be the interaction between anatomic, metabolic and vascular
networks (wires) and their coupling to functional correlations
(data) \cite{sporns}. Here instead we adopted the theory of network of
networks to investigate the optimality of coupled functional brain
modules. The consistency between experimental data and theoretical
predictions even in this broaden notion of coupled networks is
suggestive of the possible broad scope of the theory making it a
candidate to study a wider range of inter-connected
networks \cite{schneider}.


\newpage

{\bf METHODS}

{\bf Experimental analysis.} The interdependent functional brain
networks are constructed from fMRI data following the methods of
Ref. \cite{Gallos2012}. First, the Blood Oxygen Level Dependent (BOLD)
signal from each brain voxel (node) is used to construct the
functional network topology based on standard
methods \cite{eguiluz,achard} using the equal-time cross-correlation
matrix, $C_{ij}$, of the activity of pairs of voxels (see SI
Section \ref{experiments}).  

The derivation of a binary graph from a continuous connectivity matrix
relies on a threshold $T$ where the links between two nodes (voxels)
$i$ and $j$ are occupied if $T<C_{ij}$ \cite{eguiluz,Gallos2012} such
as in bond percolation. A natural and non-arbitrary choice of
threshold can be derived from a clustering bond percolation
process. The size of the largest connected component of voxels as a
function of $T$ reveals clear percolation-like
transitions \cite{Gallos2012} in the two datasets identified by the
jumps in the size of the largest component in Fig.~\ref{fig3.fig}a-b.
The emergent networks in resting state correspond to the medial prefrontal
cortex, posterior cingulate, and lateral temporoparietal regions, all
of them part of the default mode network (DMN) typically seen in
resting state data \cite{Shehzad2009}. In dual-task, as expected
for an experiment involving visual and auditory stimuli and bi-manual
responses, the responsive regions include bilateral visual
occipito-temporal cortices, bilateral auditory cortices, motor,
premotor and cerebellar cortices, and a large-scale bilateral
parieto-frontal structure.

{\bf Scaling of correlations in the brain.} We identify functional
networks (see Fig.~\ref{fig3.fig}a-b right panels) made of nodes
connected by strong links (strong BOLD signal correlation $C_{ij}$)
which are interconnected by weak links (weak BOLD signal
correlation) \cite{Gallos2012,bialek}.  Statistical analysis based on
standard maximum likelihood and KS methods \cite{clauset} (see SI
Section \ref{ks}) yield the values of the indegree exponents of each
functional brain network: $\gamma=2.85\pm 0.04$ and $k_{\rm max}=133$
for resting state and $\gamma=2.25\pm 0.07$, $k_{\rm max}=139$ for
dual-task (Fig.~\ref{fig3.fig}c-d). The obtained exponent $\alpha$
shows high positive values for both experiments: $\alpha= 1.02\pm0.02$
and $0.92\pm0.02$ for resting state and dual task data, respectively
(Fig.~\ref{fig3.fig}e). The inter-network connections show positive
exponents for both systems: $\beta=0.66\pm0.03$ and
$\beta=0.79\pm0.04$ for resting state and dual-task, respectively
(Fig.~\ref{fig3.fig}f).

Hence, in accordance with the predictions of the theory, these two
interdependent brain networks derived from qualitatively distinct
mental states (resting states and strong engagement in a task which
actively coordinates visual, auditory and motor function) show
consistently high values of $\alpha$ and positive values of $\beta$.
Figure ~\ref{fig4.fig} shows the theoretical phase diagram
$p_c(\alpha,\beta)$ in conditional and redundant mode calculated for
coupled networks with the experimental values $\gamma=2.25$ and
$2.85$.  Left panels show the prediction of $p_c(\alpha,\beta)$ in the
conditional mode of failure and right panels correspond to the
redundant mode. The experimental $(\alpha,\beta)$ are shown in white
circles lying in stable regions of the phase diagram (low $p_c$).
Interestingly, the convergence of inter-network connections, $\beta$,
is slightly higher under task conditions, adding a new degree of
freedom to the system of networks, the dynamic allocation of
functional connections governed by context-dependent processes such as
attention or learning for the case of brain networks. Further research
is assured to investigate the neuronal mechanisms underlying
inter-network communication routines specified by $\beta$.

{\bf Acknowledgements} 

This work was funded by NSF-PoLS PHY-1305476 and NIH-NIGMS
1R21GM107641. We thank N. A. M. Ara\'ujo, S. Havlin, L. Parra,
L. Gallos, A. Salles and T. Bekinschtein for clarifying
discussions. Additional financial support was provided by CNPq, CAPES,
FUNCAP, the Spanish MINECO BFU2012-39958, CONICET and the James
McDonnell Foundation 21st Century Science Initiative in Understanding
Human Cognition - Scholar Award.

{\bf Author contributions}

All authors contributed equally to the work presented in this paper.

{\bf Additional information}

The authors declare no competing financial interests. Supplementary
information accompanies this paper on
www.nature.com/naturephysics. Reprints and permissions information is
available online at
http://npg.nature.com/reprintsandpermissions. Correspondence and
requests for materials should be addressed to H.A.M.

\newpage

FIG. \ref{fig1.fig}. {\bf Modeling degree-degree correlations between
interconnected networks.}  {\bf a,} Hubs (red nodes) and non-hubs
(blue nodes) have $k_{\rm out}$ outgoing links (wiggly blue links)
according to the parameter $\alpha$. When $\alpha<0$, the outgoing
links are more likely to be found attached to non-hub nodes. When
$\alpha>0$, hubs are favored over non-hub nodes. Nodes from different
networks are connected according to $\beta$. When $\beta>0$, nodes
with similar degree prefer to connect between themselves, and when
$\beta<0$, nodes connect dissasortatively.  For simplicity we
exemplify the outgoing links emanating from only a few nodes in
network $A$ according to ($\alpha,\beta)$.
{\bf b,} Conditional mode of failure: a node fails every time it
becomes disconnected from the largest component of its own network, or
looses all its outgoing links. All stable nodes have at least one
out-going link. We exemplified only one cascading path for simplicity.
In reality, we investigate the cascading produced by removal of $1-p$
nodes from both networks. With the failure of the hub indicated in the
figure (Stage 1), all its non-hub neighbors also fail because they
become isolated from the giant component in $A$ (Stage 2). In Stage 3
the upper hub from network $B$ fails, due to the conditional
interaction, since it looses connectivity with network $A$ even though
it is still connected in $B$. With the failure of this second hub all
its non-hub neighbors become isolated, leading to their failure (Stage
4). This leads to a further removal of the second outgoing link and
the cascading failure propagates back to network $A$ (Stage 5). Since
no more nodes become isolated, the cascading failure stops with the
mutual giant component shown in Stage 5. At this point we measure the
fraction of nodes in the giant component of $A$ and $B$.  {\bf c,}
Redundant interaction: The failure of a node only leads to further
failure if its removal isolates its neighbors in the same network. The
failure of the hub (Stage 1) do not propagate the damage to the other
network (Stage 2 and 3) and therefore there is no cascading in this
interaction. We measure the fraction of nodes in the mutually
connected giant component. We note that nodes can be stable even if
they do not have out-going links as long as they belong to the
mutually connected component. Thus, the mutually connected giant
component may contain nodes which are not part of the single giant
component of one of the networks as shown in Stage 3, network $A$.

FIG. \ref{fig2.fig}. {\bf Stability phase diagram of
$p_c(\alpha,\beta)$ for conditional and redundant failure.}
Percolation threshold $p_c(\alpha,\beta)$ predicted by theory for
coupled networks for generic values $\gamma=2.5$ and $k_{\rm max}=100$
in {\bf a,} conditional interaction and {\bf b,} redundant
interaction. We use a bounded power-law for closer comparison with
experimental data. For a given system, the results are independent of
the cut-off. For the conditional interaction the system is more stable
(low value of $p_c$) when $\alpha<0$ as well as for $\alpha\approx 1$
and $\beta>0$, and displays a maximum in $p_c$ (unstable) around
$\alpha\approx 0.25$ and $\beta<0$. The redundant interaction instead
is most unstable for $\alpha<0$ and becomes stable for $\alpha\approx
1$ and $\beta>0$.  Thus the best compromise between both modes of
failures is for values located in the upper-right quadrant
$\alpha\approx 1$, $\beta>0$.

FIG. \ref{fig3.fig}. {\bf Analysis of interconnected functional brain
networks.}  {\bf a,} Clustering analysis to obtain the system of
networks for resting state data for a typical subject out of 12 scans
analyzed. Left plot shows the fraction of nodes in the largest network
versus $T$. We identify one percolation-like transition with the jump
at $T_c=0.854$. Strong in-going links define the networks and
correspond to $T>T_c$ \cite{Gallos2012}. At $T_c$, the two largest
networks, shown in the right panel in the network representation and
in the inset in the brain, merge. Interconnecting weak out-going links
are defined for $0.781<T<T_c$ (plotted in grey).
{\bf b,} The same clustering analysis is done to identify the
interconnected network in dual task \cite{Gallos2012}.  We show a
typical scan out of a total of 16 subjects.  The strong ingoing links
have $T>T_c=0.914$, and weak outgoing links $0.864<T<T_c$.
{\bf c,} The in-degree $k_{\rm in}$ distribution for the resting state
and {\bf d,} dual task experiment.  {\bf e,} Out-degree $k_{\rm
out}$ as a function of $k_{\rm in}$ for resting state and dual
task, according to Eq. (\ref{alpha.eq}).  {\bf f,} $k_{\rm in}^{\rm
nn}$ as a function of $k_{\rm in}$ for resting state and dual task
experiments, according to Eq. (\ref{beta.eq}).

FIG. \ref{fig4.fig}. {\bf Stability phase diagram for brain networks.}
Percolation threshold $p_c(\alpha,\beta)$ obtained from theory for two
coupled networks with power-law exponents and cutoff given by the
brain networks in {\bf a,} resting state and {\bf b,} dual task.  The
left panels are for conditional interactions and the right panels for
redundant interactions. The white circles represent the data points of
the real brain networks. They indicate that the brain structure
results from a compromise of optimal stability between both modes of
failure.

\newpage

\clearpage

\clearpage

\clearpage

\newpage

\begin{table*}[t]
\begin{tabular}{|| c | c | c | c | c ||}
\hline
Dataset
	& $\gamma$ & $\alpha$
	& $\beta$  & $k_{\rm max}$ \\ \hline\hline

Human Resting State
	& \pf $2.85\pm0.04$ \pf & $1.02\pm0.02$
	& \pf$0.66\pm0.03$ & 133 \\

Human Dual Task & $2.25\pm0.07$ & $0.92\pm0.02$ & \pf$0.79\pm0.04$
	& 139 \\
\hline
%

\end{tabular}
\caption{Parameters characterizing the studied human brain networks.}
\label{exp.table}
\end{table*}

\clearpage

\centerline{\bf SUPPLEMENTARY INFORMATION}

\centerline{\bf Avoiding catastrophic failure in correlated
network of networks}

\centerline{Reis, Hu, Babino, Andrade, Canals, 
Sigman, Makse}


\section{Theory of correlated network of networks}
\label{generating}

We first illustrate the theory to calculate the percolation threshold
for a single uncorrelated network following the standard calculations
done by Moore and Newman \cite{Moore2000}. We then generalize this
theory to the case of two correlated interconnected networks to
calculate $p_c$ under redundant and conditional modes of failures.

\subsection{Calculation of percolation threshold for a single network \cite{Moore2000}}
\label{single}


The percolation problem of a single network can be solved by the
calculation of the probability $X$ to reach the giant component by
following a randomly chosen link \cite{Moore2000}. First, choose a
link of a single network at random. After that, select one of its ends
with equal probability. The probability $1-X$ is the probability that,
by following this link using the chosen direction, we do not arrive at
the giant component, but instead we connect to a finite component.

Since the degree distribution of an end node of a chosen link is given
by $kP(k)/\av{k}$, one can write down a recursive equation for $X$
as:
\begin{equation}
X=1-\sum_{k}\frac{kP(k)}{\av{k}}(1-X)^{k-1}.
\end{equation}
The sum is for the probability that, by following the chosen link, we
arrive at a node with degree $k$ which is not attached to the giant
component through its remaining $k-1$ connections.  We rewrite the
previous equation as follows:
\begin{equation}
 X=1-\sum_{k}\frac{kP(k)}{\av{k}}\mathcal{G}(X),  \label{eq.singleX}
\end{equation}
where 
\begin{equation}
\label{G}
\mathcal{G}(X)=(1-X)^{k-1}.
\end{equation}

Once the probability $X$ is known, we can use it to write the probability
$1-S$ that a randomly chosen node does not belong to the giant component.
Again, this is a sum of probabilities: the probability that this node has
no links attached to it, plus the probability that this node has one link
and this link does not lead to the giant component, plus the probability
that this node has two links and none of them leads to the giant component,
and so on. In other words:
\begin{equation}
 1-S=\sum_{k}P(k)(1-X)^{k}.
\end{equation}
Again, we can rewrite this equation as:
\begin{equation}
 S=1-\sum_{k}P(k)\mathcal{H}(X),
 \label{eq.singleS}
\end{equation}
where 
\begin{equation}
\label{H}
\mathcal{H}(X)=(1-X)^{k}.
\end{equation}
Note that the probability $S$ not only stands for the probability of
choosing one node from the giant component at random, but also
provides the fraction of nodes in the network occupied by the giant
component. Equation (\ref{eq.singleS}) provides the probability of a
node to belong to the giant component and is the main quantity to be
calculated by the theory from where the value of the percolation
threshold can be calculated as the largest value of $p_c$ such that
$S(p_c)=0$.

\subsection{Analytical approach for two interconnected networks with 
correlations}
\label{correlation}

Now, we present a generalization of the above approach suited to both
problems studied in our work, namely, the redundant and conditional
interactions of two interconnected networks with generic correlations.
We have also developed an analogous theoretical framework based on the
generating approach used in Ref. \cite{Buldyrev2010}. However, we find
that the generating function approach \cite{Buldyrev2010} is more
mathematically cumbersome if one wants to take into account the
correlations between the networks to calculate the mutually connected
giant component. Since the size of the giant component is the only
quantity needed in this study, we find that the approach of Moore and
Newman is more transparent and, furthermore, allows us to take into
account both modes of failure in a single theory. Indeed, the whole
theory can be cast into a few number of equations, while the
generating function approach is more involved.

We define two probabilities for network $A$ (and their equivalents for
network $B$).  As we did for the case of a single network, we will
take advantage of functions similar to $\mathcal{G}(X)$ and
$\mathcal{H}(X)$. By doing this, the following recursive equations are
general and can be applied to the redundant and to the conditional
interaction cases depending of the way the functions
$\mathcal{G}(\bullet)$ and $\mathcal{H}(\bullet)$ are written for each
case. Therefore, below we develop the theory for both modes of failure
and later we specialize on each interaction.

First, we define the probability $X_{\rm A}$, as the probability that,
by following a randomly chosen link of network $A$, we reach a node
from the largest connected component of network $A$. The second
probability, $Y_{k_{\rm in}^{\rm A}}$, is the probability of choosing
at random a node from network $A$ with in-degree $k_{\rm in}^{\rm A}$
connected with a node from the largest component of network
$B$. Analogously, we define probabilities $X_{\rm B}$ and $Y_{k_{\rm
in}^{\rm B}}$ for network $B$.

Thus, if we initially remove a fraction $1-p_{_A}$ of nodes from
network $A$ chosen at random, and a fraction $1-p_{_B}$ of nodes from
network $B$, we can write $X_{\rm A}$ and $X_{\rm B}$ in analogy with
Eq. (\ref{eq.singleX}) [we note that when network $A$ and network $B$
have the same number of nodes, $p=(p_{_A}+p_{_B})/2$]:

\begin{widetext}
\begin{equation}
X_{\rm A} = p_{_A}\left[1-\sum_{k_{\rm in}^{\rm A},k_{\rm out}^{\rm A}}
  \frac{k_{\rm in}^{\rm A} P\left(k_{\rm in}^{\rm A},k_{\rm out}^{\rm A}\right)}{\av{k_{\rm in}^{\rm A}}}
  \mathcal{G}(X_{\rm A},Y_{k_{\rm in}^{\rm A}},k_{\rm in}^{\rm A},k_{\rm out}^{\rm A})\right].
\label{X_A.eq}
\end{equation}
\end{widetext}
Here, the correlations between $k_{\rm in}^{\rm A}$ and $k_{\rm
out}^{\rm A}$ from Eq.~(\ref{alpha.eq}) are quantified by $P(k_{\rm
in}^{\rm A},k_{\rm out}^{\rm A})$, which is the joint probability
distribution of in- and out-degrees of nodes from network $A$ from
where Eq.~(\ref{alpha.eq}) can be derived. The probability function
$\mathcal{G}(X_{\rm A},Y_{k_{\rm in}^{\rm A}},k_{\rm in}^{\rm
A},k_{\rm out}^{\rm A})$ in Eq.~(\ref{X_A.eq}) is analogous to
Eq. (\ref{G}). It stands for the probability that, by following a
randomly chosen link from network $A$, we reach a node which is not
part of the giant component of network $A$, which has in-degree
$k_{\rm in}^{\rm A}$ and out-degree $k_{\rm out}^{\rm A}$ and/or is
not connected with a node from the giant component network $B$~(here
and in what follows, ``and/or'' refers to the nature of the two cases
of study: the
redundant and conditional interactions, respectively). To write down
Eq.~(\ref{X_A.eq}) we use the joint in- and out-degree distribution of
an end node of a randomly chosen in-link $k_{\rm in}^{\rm A}P(k_{\rm
in}^{\rm A},k_{\rm out}^{\rm A})/\av{k_{\rm in}^{\rm A}}$.  Finally,
the terms in the squared brackets stand for the probability $X_{\rm
A}=X_{\rm A}(p_{_A}=1)$ before removing the fraction $1-p_{_A}$, which
is the generalization of Eq.~(\ref{eq.singleX}). Thus, after the
removal of a fraction $1-p_{_A}$, the probability of following a
randomly selected in-link to reach a node which belongs to the giant
cluster of $A$ is $X_{\rm A}(p_{_A}=1)$ times the probability $p_{_A}$
for this node being a survival node. In a similar fashion, we write
the probability $X_{B}$, the joint degree distribution $P(k_{\rm
in}^{\rm B},k_{\rm out}^{\rm B})$ and the probability function
$\mathcal{G}(X_{\rm B},Y_{k_{\rm in}^{\rm B}},k_{\rm in}^{\rm
B},k_{\rm out}^{\rm B})$ for network $B$:
%
\begin{widetext}
\begin{equation}
X_{\rm B} = p_{_B}\left[1-\sum_{k_{\rm in}^B,k_{\rm out}^{\rm B}}
  \frac{k_{\rm in}^{\rm B} P\left(k_{\rm in}^{\rm B},k_{\rm out}^{\rm B}\right)}{\av{k_{\rm in}^{\rm B}}}
  \mathcal{G}(X_{\rm B},Y_{k_{\rm in}^{\rm B}},k_{\rm in}^{\rm B},k_{\rm out}^{\rm B})\right].
\label{X_B.eq}
\end{equation}
\end{widetext}

For the probability $Y_{k_{\rm in}^{\rm A}}$ of choosing at random a
node from the network $A$ with degree $k_{\rm in}^{\rm A}$ connected
through an out-link with a node from the giant component of $B$, we
write down the following expression:
\begin{equation}
  Y_{k_{\rm in}^{\rm A}}=p_{_B}\left[1-\sum_{k_{\rm in}^{\rm B}}
      P\left(k_{\rm in}^{\rm B}|k_{\rm in}^{\rm A}\right)
      \left(1-X_{\rm B}\right)^{k_{\rm in}^{\rm B}}\right].
\label{Y_{k_s^A}.eq}
\end{equation}
The term inside the squared brackets is the probability of choosing a
node from network $B$ which is not part of the giant component of $B$
and it is connected with a node from network $A$ of in-degree $k_{\rm
in}^{\rm A}$.  Naturally, $Y_{k_{\rm in}^{\rm A}}$ is this probability
times the probability $p_{_B}$ of the $B$-node being a survival node
after the removal of a fraction $1-p_{_B}$ of nodes from network
$B$. To write down this equation, we use the conditional probability
$P(k_{\rm in}^{\rm B}|k_{\rm in}^{\rm A})$ of a node from network $B$
with in-degree $k_{\rm in}^{\rm B}$ being connected with a node with
in-degree $k_{\rm in}^{\rm A}$ from network $A$, and the probability
that, by following an in-link from $B$, we do not reach the giant
component of $B$, $(1-X_{\rm B})$. The conditional probability
$P\left(k_{\rm in}^{\rm B}|k_{\rm in}^{\rm A}\right)$ quantify the
correlations expressed by Eq.~(\ref{beta.eq}). Similar equation can be
written for $Y_{k_{\rm in}^{\rm B}}$:
%
%
\begin{equation}
 Y_{k_{\rm in}^{\rm B}}=p_{_A}\left[1-\sum_{k_{\rm in}^{\rm A}}
      P\left(k_{\rm in}^{\rm A}|k_{\rm in}^{\rm B}\right)
      \left(1-X_{\rm A}\right)^{k_{\rm in}^{\rm A}}\right].
\label{Y_{k_s^B}.eq}
\end{equation}

With $X_{\rm A}$, $X_{\rm B}$, $Y_{k_{\rm in}^{\rm A}}$, and
$Y_{k_{\rm in}^{\rm B}}$ on hand, it is possible to compute the
fraction of survival nodes in the giant component of network $A$,
$S_{\rm A}$, and in network $B$, $S_{\rm B}$, through the relations
analogous to Eq. (\ref{eq.singleS}):
\begin{widetext}
\begin{equation}
 S_{\rm A} = p_{_A}\left[1 -
    \sum_{k_{\rm in}^{\rm A},k_{\rm out}^{\rm A}}
    P(k_{\rm in}^{\rm A},k_{\rm out}^{\rm A})
    \mathcal{H}(X_{\rm A},Y_{k_{\rm in}^{\rm A}},k_{\rm in}^{\rm A},k_{\rm out}^{\rm A})\right],
\label{S_A}
\end{equation}
and
 \begin{equation}
  S_{\rm B} = p_{_A}\left[1 -
     \sum_{k_{\rm in}^{\rm B},k_{\rm out}^{\rm B}}
     P(k_{\rm in}^{\rm B},k_{\rm out}^{\rm B})
     \mathcal{H}(X_{\rm B},Y_{k_{\rm in}^{\rm B}},k_{\rm in}^{\rm B},k_{\rm out}^{\rm B})\right].
\label{S_B}
 \end{equation}
\end{widetext}

The probability function $\mathcal{H}(X_{\rm A},Y_{k_{\rm in}^{\rm
A}},k_{\rm in}^{\rm A},k_{\rm out}^{\rm A})$ generalizes
Eq. (\ref{H}), and stands for the probability of randomly selecting a
node from network $A$ with in-degree $k_{\rm in}^{\rm A}$ and
out-degree $k_{\rm out}^{\rm A}$, which is not in the giant component
of $A$ and/or it is not connected with the giant component of $B$
(again, and/or refers to redundant and conditional modes of
interaction, respectively).

\begin{figure*}
 \begin{center}
  \includegraphics[width=.9\columnwidth]{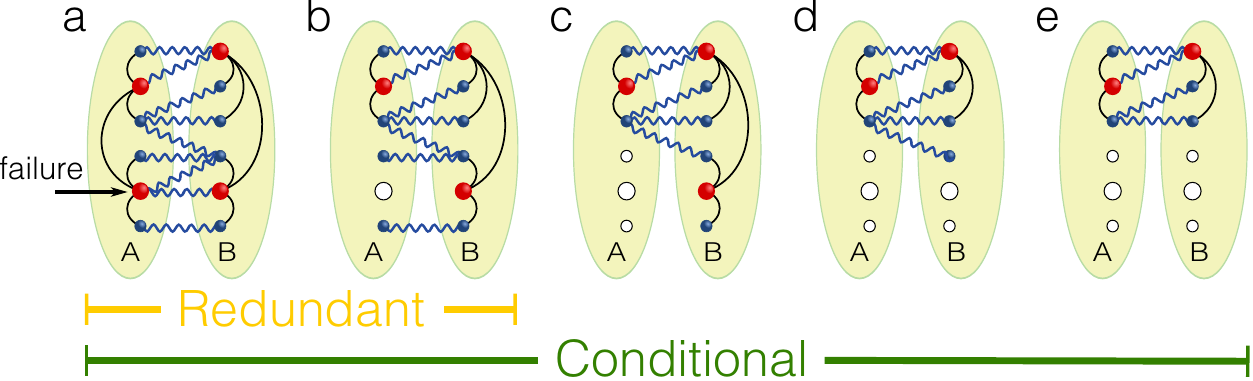}
 \end{center}
\caption{Pictorial representation of the {\bf a}-{\bf e} conditional and {\bf a}-{\bf b} redundant modes of
interaction. {\bf a,} One node is removed, or fails, in network $A$,
{\bf b,} as in a regular percolation process this node is removed
together with its links. In the redundant mode of interaction, the
neighbors of this node are not removed, because they still maintain
connection with the giant component from network $B$, but {\bf c,} for
the conditional mode of interaction the two nodes are removed, since
they do not belong to the giant component of network $A$. {\bf d,} As
a consequence of the removal of the nodes in network $A$ all the nodes
from network $B$ that lose connectivity with network $A$ are also
removed. {\bf e,} Finally, the last node from network $B$ is removed
once it loses connectivity with the giant component of network $B$. In
the end, for the conditional mode of interaction, only the mutually
connected component remains.}
\label{pictorial}
\end{figure*}

Due to the different meanings that the probability function
$\mathcal{H}(X_{\rm A},Y_{k_{\rm in}^{\rm B}},k_{\rm in}^{\rm A},
k_{\rm out}^{\rm A})$ may assume depending of the mode of interaction,
for this general approach the nature of the quantities $S_A$ and $S_B$
differ conceptually from the quantity $S$ presented by Eq.
(\ref{eq.singleS}) for a single network. See Fig. \ref{pictorial} for
more details. For the conditional mode, a node, or a set of nodes from
network $A$, for example, will fail if (i) it loses connection with
the largest component of network $A$, or if (ii) it loses connection
with the largest component of network $B$.  Thus $\mathcal{H}(X_{\rm
A},Y_{k_{\rm in}^{\rm B}},k_{\rm in}^{\rm A}, k_{\rm out}^{\rm A})$ is
the probability function that describes the probability of picking a
node at random from network A that is not part of the largest
component of A (due to condition (i) this node will fail) {\bf or}
that is not connected to the largest cluster of network B (due to
condition (ii) this node will also fail). Thus, $S_A$ (and its
counterpart $S_B$ for network $B$) is the fraction occupied by the
largest component of survival node in network $A$. For a finite size
network, $S_A=n_A/N_A$, where $n_A$ is the number of nodes in the
largest component and $N_A$ the number of nodes in network $A$. It is
important to note that due to the condition (ii) this fraction is
necessarily the same as the size of the giant connected component of
network $A$. $S_A$ may be interpreted also as the fraction from
network $A$ that is part of the mutually connected giant component
$S_{\rm AB}$, as in Ref.~\cite{Buldyrev2010}.  The same applies to
network $B$. In other words, the number of nodes in the mutually
connected giant component belonging to $B$ is the same as the number
of nodes in the giant connected component of $B$ as calculated after
the attack as if $B$ was a single network.

For the redundant mode, since there is no cascading propagation of
damage due to the failure of a neighbor, $\mathcal{H}(X_{\rm
A},Y_{k_{\rm in}^{\rm B}},k_{\rm in}^{\rm A}, k_{\rm out}^{\rm A})$ is
the function that describes the probability of picking a node at
random, for example from network $A$, which is not connected to the
largest component from its own network, network $A$, {\bf and} is not
connected to the largest component of network $B$ via an out-going
link. Therefore, the quantity $S_A$ provides the fraction of
``active'' nodes, or in other words, the fraction of survival nodes
that may be part of the largest component of network $A$, and in
addition a fraction from network $A$ that are disconnected from that
largest component of network $A$, but are not failed because they are
still connected to the largest component of network $B$ via an
out-going link. Thus, the mutually connected giant component $S_{\rm
AB}$ has a different structure in this mode compared to the
conditional mode. This situation is illustrated in
Fig. \ref{fig1.fig}c and \ref{pictorial}. At the end of the attack
process, there is a remaining node in network $A$ which is not
connected to the giant component of $A$ calculated as if it is a
single network. Such a node is still ``on'' since it is connected to
$B$ via an out-going link. Thus, the mutually connected giant
component contains this node.

Furthermore, a node that has lost all its out-going link will fail in
the conditional interaction, even if it is still connected to its own
giant component. However, in the redundant mode, a node without
out-going links may still function as long as it is still connected to
the giant component of its own single network. For instance, many
nodes are still functioning in Fig. \ref{fig1.fig}c, redundant mode,
even though they are not interconnected. However, in conditional
interaction Fig. \ref{fig1.fig}b, all stable nodes needs to have
out-going links.  That is, in redundant mode, the nodes can still
receive power via the same network or the other network, while in the
conditional node, they need out-going connectivity all the
time. Taking into account these considerations, the value of $p_c$ is
obtained from the behavior of the giant component of either of the
networks in the conditional mode, while in the redundant mode, the
value of $p_c$ is obtained from the size of the mutually giant
connected component. However, in this last case, it is statistically
the same to obtain $p_c$ from the giant components of one of the
networks as well. In what follows the calculations of the giant
components are done by considering two networks of equal size $N$ and
damaging each network with a fraction $1-p$ of nodes.

Next, we explicitly write the probability functions $\mathcal{G}$
and $\mathcal{H}$ for both, conditional and redundant interactions,
respectively, to occur on interactive networks after a random failure
of $1-p_{_A}$ and $1-p_{_B}$ nodes.
It is important to note that the probabilities $\mathcal{G}$ and
$\mathcal{H}$ describe the probability of randomly choosing a node
which is not part of the giant component of one network and/or is not
connected to a node from the giant component of the adjacent
network. In other words, this node picked at random is not part of the
giant component of the whole network.  We test the general case where
both networks are attacked: $p_{_A}\neq1$ and $p_{_B}\neq1$. The theory
can be used to attacking only one network by setting $p_{_B}=1$.

{\bf Redundant interaction: } 
We consider the total fraction $1-p$ of nodes removed from the two
networks. If network $A$ and network $B$ have the same number of
nodes, then $p=(p_{_A}+p_{_B})/2$.  For redundant interaction two
events are important.  Both events are defined as follows. The first
is the probability that, by following a randomly chosen link of a
network, we do not reach the giant component of that network. For
network $A$, this probability can be written as $(1-X_{\rm A})$. The
second is the probability of choosing at random a node from one
network, say network $A$, with in-degree $k_{\rm in}^{\rm A}$ which is
not connected with a node from the giant component of network
$B$. This probability can be written as $(1-Y_{k_{\rm in}^{\rm A}})$.
In the case of redundant interaction (with no cascading due to
conditional mode) these two probabilities are independent, since the
lack of connectivity with network $B$ does not imply failure of a node
from network $A$. Thus, the probability function $\mathcal{G}(X_{\rm
A},Y_{k_{\rm in}^{\rm A}},k_{\rm in}^{\rm A},k_{\rm out}^{\rm A})$
that, by following a randomly selected link we arrive at a node with
in-degree $k_{\rm in}^{\rm A}$ and out-degree $k_{\rm out}^{\rm A}$
which is not part of the giant cluster of its own network and is not
connected with a node from the giant cluster of the adjacent network
can be written as:
\begin{equation}
 \mathcal{G}(X_{\rm A},Y_{k_{\rm in}^{\rm A}},k_{\rm in}^{\rm A},k_{\rm out}^{\rm A})
	=(1-X_{\rm A})^{k_{\rm in}^{\rm A}-1}(1-Y_{k_{\rm in}^{\rm A}})^{k_{\rm out}^{\rm A}}.
\label{eq.GA_red}
\end{equation}
%

Similarly, the probability function
$\mathcal{H}(X_{\rm A},Y_{k_{\rm in}^{\rm A}},k_{\rm in}^{\rm A},k_{\rm out}^{\rm A})$
of picking a node, at random, with in-degree $k_{\rm in}^{\rm A}$ and
out-degree $k_{\rm out}^{\rm A}$ from one network which is not part of
the giant cluster of its own network and is not connected with a node
from the giant cluster from the adjacent network is:
\begin{equation}
 \mathcal{H}(X_{\rm A},Y_{k_{\rm in}^{\rm A}},k_{\rm in}^{\rm A},k_{\rm out}^{\rm A})
	=(1-X_{\rm A})^{k_{\rm in}^{\rm A}}(1-Y_{k_{\rm in}^{\rm A}})^{k_{\rm out}^{\rm A}}.
\label{eq.HA_red}
\end{equation}
%
Again, we can write equivalent expressions for
$\mathcal{G}(X_{\rm B},Y_{k_{\rm in}^{\rm B}},k_{\rm in}^{\rm B},k_{\rm out}^{\rm B})$
and
$\mathcal{H}(X_{\rm B},Y_{k_{\rm in}^{\rm B}},k_{\rm in}^{\rm B},k_{\rm out}^{\rm B})$
as
\begin{equation}
 \mathcal{G}(X_{\rm B},Y_{k_{\rm in}^{\rm B}},k_{\rm in}^{\rm B},k_{\rm out}^{\rm B})
	=(1-X_{\rm B})^{k_{\rm in}^{\rm B}-1}(1-Y_{k_{\rm in}^{\rm B}})^{k_{\rm out}^{\rm B}},
\label{eq.GB_red}
\end{equation}
and
\begin{equation}
 \mathcal{H}(X_{\rm B},Y_{k_{\rm in}^{\rm B}},k_{\rm in}^{\rm B},k_{\rm out}^{\rm B})
	=(1-X_{\rm B})^{k_{\rm in}^{\rm B}}(1-Y_{k_{\rm in}^{\rm B}})^{k_{\rm out}^{\rm B}}.
\label{eq.HB_red}
\end{equation}

{\bf Conditional interaction: } This interaction leads to cascading
processes.  In the conditional interaction process, we are interested
in the cascading effects on the coupled networks, $A$ and $B$, due to
an initial random failure of a portion of nodes in both networks,
where $p_{_A}\neq1$ and $p_{_B}\neq1$.  In the case of attacking
network $A$ only, the fraction $p_{_B}$ is set to be equal to one,
such that a node from network $B$ can only fail due to the conditional
interaction.

For the conditional interaction, $\mathcal{G}(X_{\rm A},Y_{k_{\rm
in}^{\rm A}},k_{\rm in}^{\rm A},k_{\rm out}^{\rm A})$ depends on the
probability that, by following a link from network $A$, we do not
arrive at a node with in-degree $k_{\rm in}$ connected to the giant
component of its own network, $(1-X_{\rm A})^{k_{\rm in}-1}$, and on
the probability of randomly choosing a node from network $A$ with
$k_{\rm out}$ outgoing links towards network $B$, $(1-Y_{k_{\rm
in}^{\rm A}})^{k_{\rm out}}$.  Also, we have the probability
$\mathcal{H}(X_{\rm A},Y_{k_{\rm in}^{\rm A}},k_{\rm in}^{\rm
A},k_{\rm out}^{\rm A})$ of picking up a node from one network which
is not part of the giant component of its own network or picking up
one node from one network which is not connected with one node from
the giant component of the adjacent network, which is also dependent
of the probabilities $(1-X_{\rm A})$ and $(1-Y_{k_{\rm in}^{\rm A}})$.

Different from the redundant mode, these probabilities, $(1-X_{\rm
A})$ and $(1-Y_{k_{\rm in}^{\rm A}})$, are not mutually exclusive in
the conditional interaction. Thus:
\begin{widetext}
\begin{equation}
     \mathcal{G}(X_{\rm A},Y_{k_{\rm in}^{\rm A}},k_{\rm in}^{\rm A},k_{\rm out}^{\rm A})
	=(1-X_{\rm A})^{k_{\rm in}^{\rm A}-1} + (1-Y_{k_{\rm in}^{\rm A}})^{k_{\rm out}^{\rm A}}
	- (1-X_{\rm A})^{k_{\rm in}^{\rm A}-1}(1-Y_{k_{\rm in}^{\rm A}})^{k_{\rm out}^{\rm A}}
	+\delta_{k_{out}^A,0}[(1-X_A)^{k_{in}^{A}-1}-1],
\label{eq.GA_cond}
\end{equation}
and
\begin{equation}
\mathcal{H}(X_{\rm A},Y_{k_{\rm in}^{\rm A}},k_{\rm in}^{\rm A},k_{\rm out}^{\rm A})
	=(1-X_{\rm A})^{k_{\rm in}^{\rm A}} + (1-Y_{k_{\rm in}^{\rm A}})^{k_{\rm out}^{\rm A}}
	 - (1-X_{\rm A})^{k_{\rm in}^{\rm A}}(1-Y_{k_{\rm in}^{\rm A}})^{k_{\rm out}^{\rm A}}
	 +\delta_{k_{out}^A,0} [(1-X_A)^{k_{in}^{A}}-1].
\label{eq.HA_cond}
\end{equation}
\end{widetext}
We can write the equivalent expressions for $\mathcal{G}(X_{\rm
B},Y_{k_{\rm in}^{\rm B}},k_{\rm in}^{\rm B},k_{\rm out}^{\rm B})$ and
$\mathcal{H}(X_{\rm B},Y_{k_{\rm in}^{\rm B}},k_{\rm in}^{\rm
B},k_{\rm out}^{\rm B})$ as follows:
\begin{equation}
\mathcal{G}(X_{\rm B},Y_{k_{\rm in}^{\rm B}},k_{\rm in}^{\rm B},k_{\rm out}^{\rm B})
	=(1-X_{\rm B})^{k_{\rm in}^{\rm B}-1} + (1-Y_{k_{\rm in}^{\rm B}})^{k_{\rm out}^{\rm B}}
	 - (1-X_{\rm B})^{k_{\rm in}^{\rm B}-1}(1-Y_{k_{\rm in}^{\rm B}})^{k_{\rm out}^{\rm B}}
	 +\delta_{k_{out}^B,0}[(1-X_B)^{k_{in}^{B}-1}-1],
\label{eq.GB_cond}
\end{equation}
and
\begin{equation}
\mathcal{H}(X_{\rm B},Y_{k_{\rm in}^{\rm B}},k_{\rm in}^{\rm B},k_{\rm out}^{\rm B})
	=(1-X_{\rm B})^{k_{\rm in}^{\rm B}} + (1-Y_{k_{\rm in}^{\rm B}})^{k_{\rm out}^{\rm B}}  
	 - (1-X_{\rm B})^{k_{\rm in}^{\rm B}}(1-Y_{k_{\rm in}^{\rm B}})^{k_{\rm out}^{\rm B}}
	 +\delta_{k_{out}^B,0} [(1-X_B)^{k_{in}^{B}}-1].
\label{eq.HB_cond}
\end{equation}
Where $\delta_{i,j}$ is the Kronecker delta.


With the set of equations (\ref{eq.GA_red})-(\ref{eq.HA_red}) and
(\ref{eq.GA_cond})-(\ref{eq.HA_cond}), and their equivalents for
network $B$, Eq.~(\ref{eq.GB_red})-(\ref{eq.HB_red}) and
(\ref{eq.GB_cond})-(\ref{eq.HB_cond}), it is possible to solve both
problems, the redundant and the conditional interactions, on a system
of two coupled networks interconnected through degree-degree
correlated outgoing nodes. The correlation between the coupled
networks is represented by the in- out-degree distribution $P(k_{\rm
in}^{\rm A},k_{\rm out}^{\rm A})$ and by the conditional probability
$P\left(k_{\rm in}^{\rm B}|k_{\rm in}^{\rm A}\right)$.  In the
following section, we present the network model used to generate a
system of two networks interconnected with correlations described by
power law functions with the exponents $\alpha$ and $\beta$. These
networks are used on the calculations of the distribution $P(k_{\rm
in}^{\rm A},k_{\rm out}^{\rm A})$ and $P\left(k_{\rm in}^{\rm
B}|k_{\rm in}^{\rm A}\right)$ for each pair of $(\alpha,\beta)$. The
final result is the probability for a node to belong to the giant
component of network $A$ or $B$-- as given by Eq. (\ref{S_A}) and
(\ref{S_B})-- as a function of the fraction of removed nodes $1-p$
(with $p_A=p_B=p$) from where the percolation threshold $p_c$ can be
evaluated from $S_A(p_c)=0$ and $S_B(p_c)=0$ as a function of the
three exponents defining the networks: $\gamma$, $\alpha$ and $\beta$,
and the cutoff in the degree distribution $k_{\rm max}$.  We use two
networks of equal size $N=1500$ nodes, each.

%

%

\subsection{Network model. Test of theory}
\label{model}

In order to test the percolation theory using the above formalism, we
need to generate a system of interacting networks with the prescribed
set of exponents and degree cutoff. The first step of our network
model is to generate two networks, $A$ and $B$, with the same number
$N$ of nodes and with the desired in-degree distribution $P(k_{\rm
in})$ as defined by $\gamma$ and the maximum degree $k_{\rm max}$.  To
do this we use the standard ``configuration model'' which has been
extensively used to generate different network topologies with
arbitrary degree distribution ~\cite{Book}.  The algorithm of
the configuration model basically consists of assigning a randomly
chosen degree sequence to the $N$ nodes of the networks in such a way
that this sequence is distributed as $P(k_{\rm in}) \sim k_{\rm
in}^{-\gamma}$ with $1 \le k_{\rm in} \le k_{\rm max}$ and $P(k_{\rm in}) =
0$ for $k_{\rm in} > k_{\rm max}$.  After that, we select a pair of
nodes at random, both with $k_{\rm in}>0$, and we connect them.

The next step of the model is to connect networks $A$ and $B$ in such
a way that their outgoing nodes have degree-degree correlations that
can be described by the parameters $\alpha$ and $\beta$ as defined in
Eqs. (\ref{alpha.eq}) and (\ref{beta.eq}). In order to do this, we use
an algorithm inspired by the configuration model. First, we assign a
sequence of out-degrees $k_{\rm out}$ to the nodes of each
network. This process is performed independently to each network by
adding the same number of outgoing links. Each outgoing link is added
individually to nodes chosen at random with a probability that is
proportional to $k_{\rm in}^{\alpha}$.  Thus, an out-degree sequence
is assigned to the nodes in each network in such a way that $k_{\rm
out}\sim k_{\rm in}^{\alpha}$ according to Eq. (\ref{alpha.eq}). This
process results in a set of outgoing stubs attached to every node in
network $A$ and $B$. The next step is to join these stubs in such a
way that we satisfy the correlations given by Eq. (\ref{beta.eq}).


The next step is to choose two nodes, one from each network, such that
$\av{k_{\rm in}^{\rm nn}}=A\times k_{\rm in}^{\beta}$, and then, we
connect them if they have available outgoing links.  Here, we choose
the factor $A$ such that $\av{k_{\rm in}^{\rm nn}}=1$ for $k_{\rm
in}=1$ when $\beta=1$, and $\av{k_{\rm in}^{\rm nn}}=k_{\rm max}$ for
$k_{\rm in}=1$ and $\beta=-1$. Thus, we write the value of the factor
as $A=A(k_{\rm max},\beta)=k_{\rm max}^{(1-\beta)/2}$.

The algorithm works as follows: we randomly choose one node $i$ from
one network. After that, we choose another node $j$, from the second
network, with in-degree $k_{\rm in}^{j}$ with probability that follows
a Poisson distribution $P(k_{\rm in}^{j}, \lambda)$, where the mean
value $\lambda=\av{k_{\rm in}^{\rm nn}}$.  We connect nodes $i$ and
$j$ if they are not connected yet.

It should be noted that Eqs. (\ref{alpha.eq}) and (\ref{beta.eq}) may
not be self-consistent for all values of $\alpha, \beta$.  For
instance, for very low values of $\beta$, e.g., $\beta=-1$, the degree
correlations between coupled networks are not always self-consistent
with the structural relations between $k_{\rm in}$ and $k_{\rm out}$
described by $\alpha$. Since $\beta$ measures the convergence of
connections between networks, when $\beta$ is negative hubs prefer to
connect with low-degree nodes. To better understand these features,
consider $\beta=-1$, and for nodes with $k_{\rm in}=1$ and $k_{\rm
in}=k_{\rm max}$. With this configuration, nodes with $k_{\rm in}=1$
are likely to be connected with nodes from the adjacent network with
$k_{\rm in}=k_{\rm max}$. When $\alpha=1$, most of the links are
attached to the highly active nodes, notably, nodes with $k_{\rm
in}=k_{\rm max}$, and less likely to nodes with $k_{\rm in}=1$. In
this regime, there are not enough low-degree nodes with outgoing links
to be connected with the high-degree nodes, thus the desired relation
between $k^{\rm nn}_{\rm in}$ versus $k_{\rm in}$ cannot be
realized. The other possible situation is when $\alpha$ is
negative. In this regime, most of the outgoing links are attached to
low-degree nodes, consequently, the few hubs from the network are
unlikely to receive an outgoing link, and even when it happens, one
hub does not have enough outgoing links to be connected to the stubs
of the low-degree nodes. For these reasons we limit our study to
$\alpha>-1$ and $\beta>-0.5$ where the relations are found to be
self-consistent.

For every initial pair $(\alpha,\beta)$, we generate a network with
the above algorithm and then we recalculate the effective values of
$(\alpha,\beta)$ which are then used to plot the phase diagram
$p_c(\alpha,\beta)$ in Fig. \ref{fig2.fig} and \ref{fig4.fig}.

\subsection{Calculation of the giant components and percolation threshold $p_c(\gamma,\alpha,\beta,k_{\rm max})$}

With the networks generated in the previous section we are able to
compute the functions $P(k_{\rm in}^{\rm A},k_{\rm out}^{\rm A})$ and
$P\left(k_{\rm in}^{\rm B}|k_{\rm in}^{\rm A}\right)$. Then we apply
the recursive equations derived previously to calculate the size of
the giant components $S_{\rm A}$ and $S_{\rm B}$ from Eqs. (\ref{S_A})
and (\ref{S_B}). We do this calculation for different values of $p$
for cases of study and then extract the percolation threshold $p_c$ at
which the giant components $S_{A}$ and $S_{B}$ vanish in conditional
mode.

\begin{figure*}
 \begin{center}
  \includegraphics[width=.65\columnwidth]{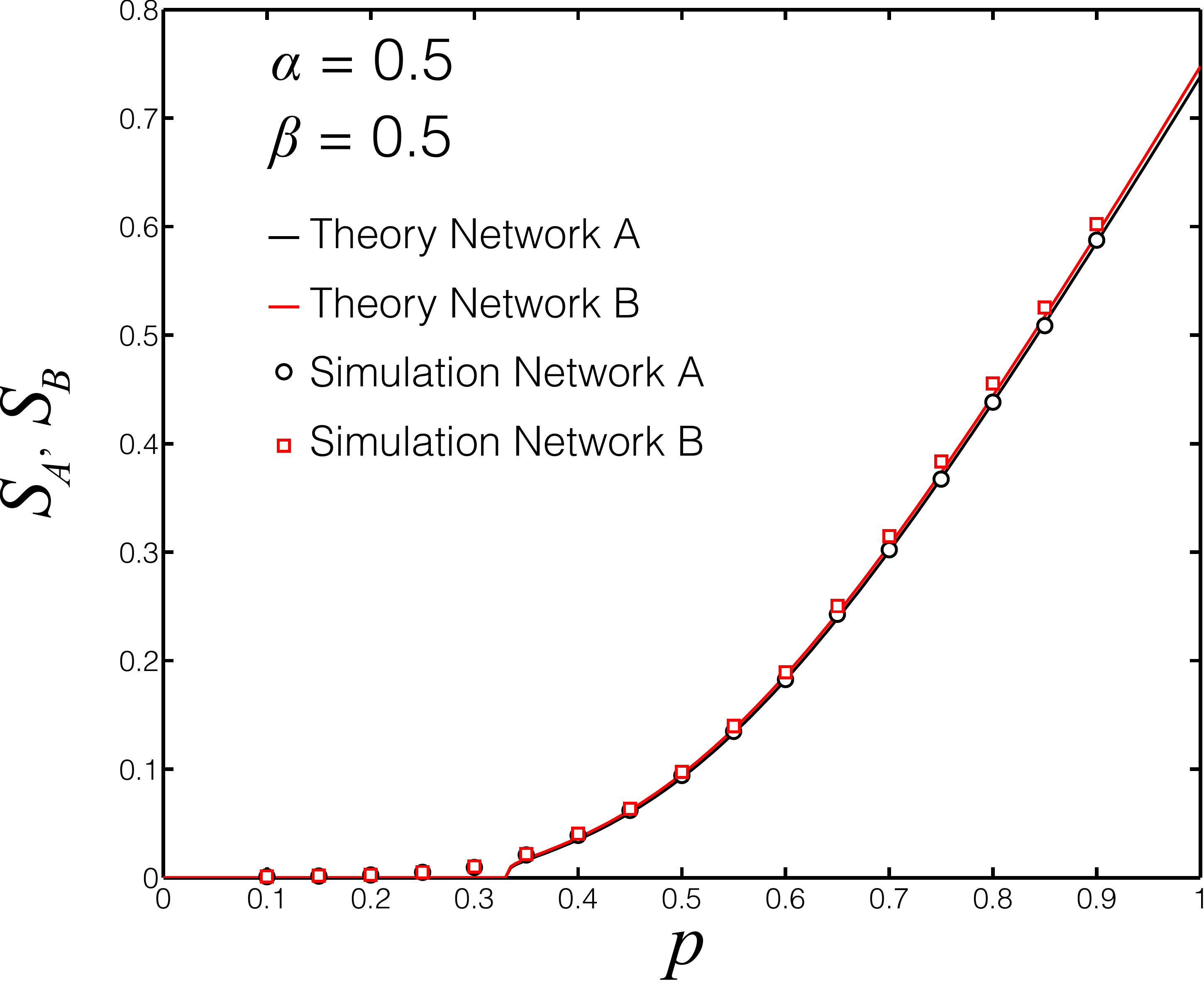}
 \end{center}
\caption{Giant component of network $A$ and $B$ in the conditional mode of
failure.  We present the prediction of the theory for values of
$N_A=N_B=1500$, $\gamma=2.5$, $\alpha=0.5$, $\beta=0.5$ and $k_{\rm
max}=100$ and compare with computer simulations of the giant component
obtained numerically by attacking the same network. We perform average
over 100 different realizations. We attack a fraction $1-p$ of both
networks and calculate the fraction of nodes belonging to the
corresponding giant components. The results show a very good agreement
between theory and simulations.}
\label{SI_giant}
\end{figure*}

Figure \ref{SI_giant} shows the predictions of the theory in the
conditional mode for a network with $\gamma=2.5$, $\alpha=0.5$, $\beta
= 0.5$ and $k_{\rm max}=100$. We plot the relative size of the giant
components in $A$ and $B$, $S_{\rm A}$ and $S_{\rm B}$, as predicted by
Eqs. (\ref{S_A}) and (\ref{S_B}).  As one can see in
Fig. \ref{SI_giant}, there is a well-defined critical value at which
the $A$-giant component vanishes which defines the percolation
threshold $p_c(\gamma,\alpha,\beta,k_{\rm max})=0.335$ for these
particular parameters.

Figure \ref{SI_giant} also presents the comparison between theoretical
results and direct simulations. We test the theory by attacking
randomly the generated correlated networks and calculating numerically
the giant components versus the fraction of removed nodes $1-p$.  The
results show a good agreement corroborating the theory.


After testing the theory, a full analysis is done spanning a large
parameter space by changing the four parameters defining the theory:
$(\gamma, \alpha, \beta, k_{\rm max})$. The results are plotted in the
main text Fig. \ref{fig2.fig} and \ref{fig4.fig} for the stated values
of the parameters. Beyond the calculation of $p_c(\alpha,\beta)$, we
also identify regimes of first-order phase transitions in the
conditional interaction, found specially when $p_c$ is high, beyond
the standard second-order percolation transition; a result that will
be expanded in subsequent papers.

\section{Experiments: Analysis of interconnected brain networks}
\label{experiments}

Our functional brain networks are based on functional magnetic
resonance imaging (fMRI).  The fMRI data consists of temporal series,
known as the blood oxygen level-dependent (BOLD) signals, from
different brain regions. The brain regions are represented by voxels.
In this work we use data sets gathered in two different and
independent experiments. The first is the NYU public data set from
resting state humans participants.  The NYU CSC TestRetest resource is
available at ~\url{http://www.nitrc.org/projects/nyu_trt/}. The second
data set was gathered in a dual-task experiment on humans previously
produced by our group
\cite{Sigman2008} and recently analyzed in Ref. \cite{Gallos2012}.
The brain networks analyzed here can be found
at:~\url{http://lev.ccny.cuny.edu/~hmakse/soft_data.html}.  Both
datasets were collected in healthy volunteers and using 3.0T MRI
systems equipped with echoplanar imaging (EPI). The first study was
approved by the institutional review boards of the New York University
School of Medicine and New York University. The second study is part
of a larger neuroimaging research program headed by Denis Le Bihan and
approved by the Comit\'e Consultatif pour la Protection des Personnes
dans la Recherche Biom\'edicale, H\^{o}pital de Bic\^{e}tre (Le
Kremlin-Bic\^{e}tre, France).

{\bf Resting state experiments:} A total of 12 right-handed
participants were included (8 women and 4 men, mean age 27,
ranging from 21 to 49). During the scan, participants were instructed
to rest with their eyes open while the word ‘‘Relax’’ was centrally
projected in white, against a black background.  A total of 197 brain
volumes were acquired. For fMRI a gradient echo (GE) EPI was
used with the following parameters: repetition time (TR) = 2.0 s; echo
time (TE) = 25 ms; angle = 90$^{\circ}$; field of view (FOV) = 192
$\times$ 192 mm; matrix = 64 $\times$ 64; 39 slices 3 mm thick. For
spatial normalization and localization, a high-resolution T1-weighted
anatomical image was also acquired using a magnetization prepared
gradient echo sequence (MP-RAGE, TR = 2500 ms; TE = 4.35 ms; inversion
time (TI) = 900 ms; flip angle = 8$^{\circ}$; FOV = 256 mm; 176
slices). Data were processed using both AFNI
(version
AFNI\_2011\_12\_21\_1014, \url{http://afni.nimh.nih.gov/afni}) and FSL
(version 5.0, \url{www.fmrib.ox.ac.uk}) and the help of
the \url{www.nitrc.org/projects/fcon_1000} batch scripts for
preprocessing.  The preprocessing consisted on: motion correcting
(AFNI) using Fourier interpolation, spatial smoothing (fsl) with
gaussian kernel (FWHM=6mm), mean intensity normalization (fsl), FFT
band-pass filtering (AFNI) with 0.08Hz and 0.01Hz bounds, linear and
quadratic trends removing, transformation into MIN152 space (fsl) with
a 12 degrees of freedom affin transformation, (AFNI) and extraction of
global, white matter and cerebrospinal fluid nuisance signals.

{\bf Dual task experiments:} Sixteen participants (7 women and 9 men,
mean age, 23, ranging from 20 to 28) were asked to perform two
consecutive tasks with the instruction of providing fast and accurate
responses to each of them. The first task was a visual task of
comparing a given number (target T1) to a fixed reference, and,
second, an auditory task of judging the pitch of an auditory tone
(target T2) \cite{Sigman2008}. The two stimuli are presented with a
stimulus onset asynchrony (SOA) varying from: 0, 300, 900 and 1200 ms.
Subjects had to respond with a key press using right and left hands,
whether the number flashed on the screen or the tone were above or
below a target number or frequency, respectively. Full details and
preliminary statistical analysis of this experiment have been reported
elsewhere \cite{Sigman2008,Gallos2012}.

Subjects performed a total of 160 trials (40 for each SOA value) with
a 12 s inter-trial interval in five blocks of 384 s with a resting
time of $\sim$ 5 min between blocks.  In our analysis we use all
scans, that is, scans coming from all SOA.  Since each of the 16
subjects perform four SOA experiments, we have a total of 64 brain
scans.
The experiments were performed on a 3T fMRI system
(Bruker). Functional images were obtained with a T2*-weighted gradient
echoplanar imaging sequence [repetition time (TR) 1.5 s; echo time 40
ms; angle 90°; field of view (FOV) 192 $\times$ 256 mm; matrix 64
$\times$ 64]. The whole brain was acquired in 24 slices with a slice
thickness of 5 mm. Volumes were realigned using the first volume as
reference, corrected for slice acquisition timing differences,
normalized to the standard template of the Montreal Neurological
Institute (MNI) using a 12 degree affine transformation, and spatially
smoothed (FWHM = 6mm).  High-resolution images (three-dimensional GE
inversion-recovery sequence, TI = 700 mm; FOV = 192 $\times$ 256
$\times$ 256 mm; matrix = 256 $\times$ 128 $\times$ 256; slice
thickness = 1 mm) were also acquired.
We computed the phase and amplitude of the hemodynamic response of
each trial as explained in M. Sigman, A. Jobert, S. Dehaene, Parsing a
sequence of brain activations of psychological times using fMRI. {\it
Neuroimage} {\bf 35,} 655-668 (2007).  We note that the present
data contains a standard preprocessing spatial smoothing with gaussian
kernel (FWHM=6mm), which was not applied in
Ref. \cite{Gallos2012}. Such smoothing produces smaller percolation
thresholds as compared with those obtained in Ref. \cite{Gallos2012}.


{\bf Construction of brain networks:} In order to build brain networks
in both experiments, we follow standard procedures in the
literature \cite{eguiluz,achard,Gallos2012}.  We first compute the
correlations $C_{ij}$ between the BOLD signals of any pair of voxels
$i$ and $j$ from the fMRI images. Each element of the resulting matrix
has value on the range $-1\leq C_{ij}\leq 1$.  If one considers that
each voxel represents a node from the brain network in question, it is
possible to assume that the correlations $C_{ij}$ are proportional to
the probability of nodes $i$ and $j$ being functionally
connected. Therefore, one can define a threshold $T$, such that if
$T<C_{ij}$ the nodes $i$ and $j$ are connected. We begin to add the
links from higher values to lower values of $T$. This growing process
can be compared to the bond percolation process.  As we lower the
value of $T$, different clusters of connected nodes appear, and as the
threshold $T$ approaches a critical value of $T_c$, multiple
components merge forming a giant component.

In random networks, the size of the largest component increases
rapidly and continuously through a critical phase transition at $T_c$,
in which a single incipient cluster dominates and spans over the
system~\cite{bollobas}. Instead, since the connections in brain
networks are highly correlated rather than random, the size of the
largest component increases progressively with a series of sharp
jumps. These jumps have been previously reported in
Ref. ~\cite{Gallos2012}.  This process reveals the multiplicity of
percolation transitions: percolating networks subsequently merge in
each discrete transition as $T$ decreases further. We observe this
structure in the two datasets investigated in this study: for the
human resting sate in Fig. \ref{fig3.fig}a and for the human dual task
in Fig.~\ref{fig3.fig}b.

\begin{figure*}
 \begin{center}
  \includegraphics[width=.65\columnwidth]{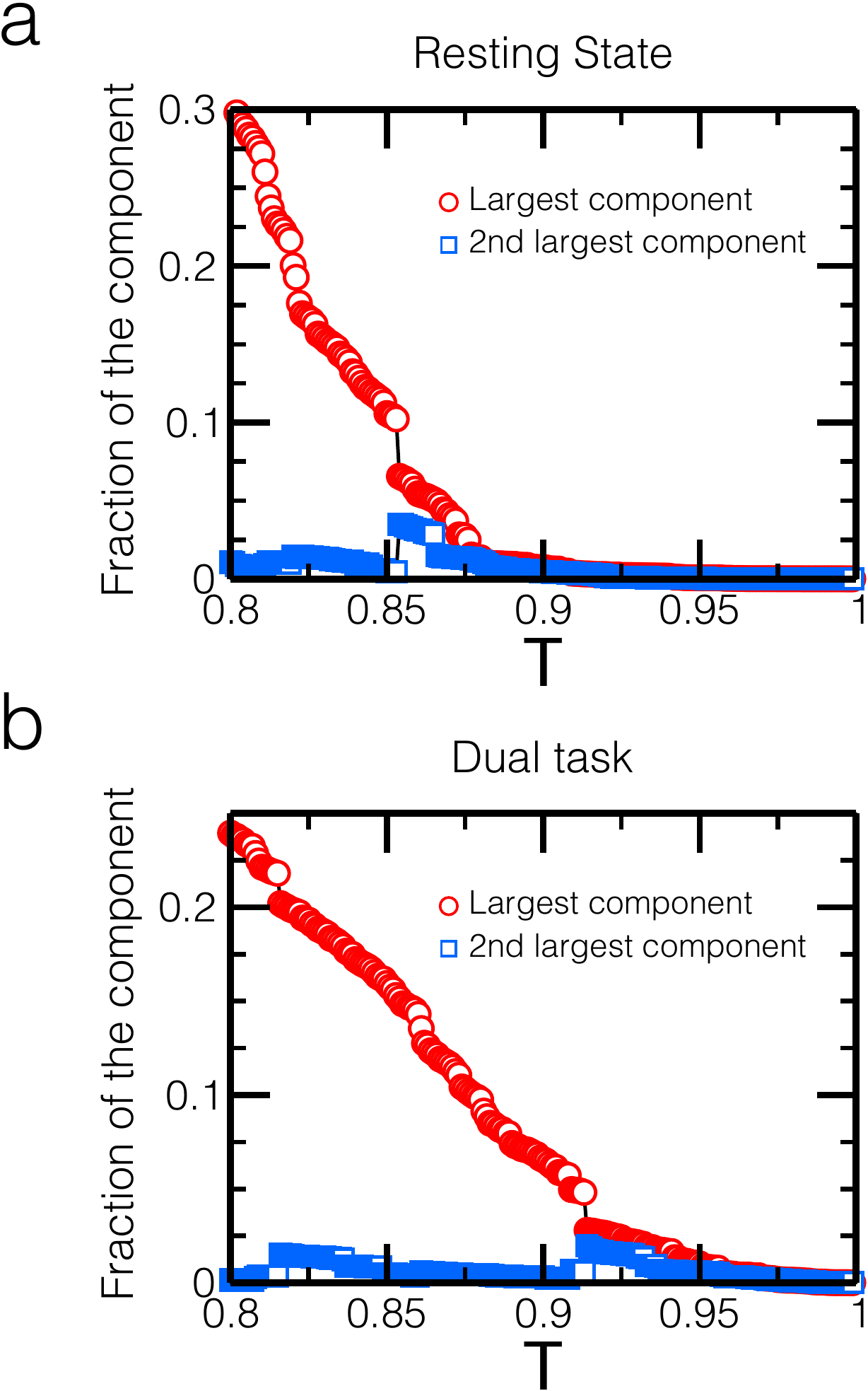}
 \end{center}
\caption{First and second largest component in the brain networks corresponding
to resting state and dual task. The largest component shows a jump
while the second largest component shows a peak, indicating a
percolation transition at $T_c$. {\bf a,} Resting state. {\bf b,} Dual
task.}
\label{second}
\end{figure*}

For each dataset we identify the critical value of $T$, namely $T_c$,
in which the two largest components merge, as one can notice in
Fig.~\ref{fig3.fig} in the main text.  While the anatomical projection
of the largest component varied across experiments, this merging
pattern at $T_c$ was clearly observed in each participant of the two
experiments analyzed here, two examples are shown in
Figs. \ref{fig3.fig}a-b. The transition is confirmed by the
measurement of the second largest cluster which shows a peak at $T_c$,
see Fig. \ref{second}.

For $T$ values larger than $T_c$ the two largest brain clusters are
disconnected, forming two independent networks. Each network is
internally connected by a set of strong-links, which correspond to
$k_{\rm in}$ \cite{Gallos2012} in the notation of systems of
networks. By lowering $T$ to values smaller than $T_c$, the two
networks connect by a set of weak-links, which correspond to $k_{\rm
out}$~\cite{Gallos2012}, i.e. the set of links connecting the two
networks.

Our analysis of the structural organization of weak links connecting
different clusters is performed with $T_0<T<T_c$. Here, $T_0$ is
chosen in such a way that the average $\av{k_{\rm out}}$ of outgoing
degrees of the nodes on the two largest clusters is $\av{k_{\rm out}}
= 1$.  For lower values of $T_0$, where $\av{k_{\rm out}} = 2$ and $=
5$, we found no relevant difference with the studied case of
$\av{k_{\rm out}} = 1$.

As done in previous network experiments based on the dual task data
\cite{Gallos2012} we create a mask where we keep voxels which were
activated in more than 75\% of the cases, i.e., in at least 48
instances out of the 64 total cases considered.  The obtained number
of activated voxels in the whole brain is $N\approx 60,000$, varying
slightly for different individuals and stimuli. The `activated or
functional map' exhibits phases consistently falling within the
expected response latency for a task-induced
activation \cite{Sigman2008}. As expected for an experiment involving
visual and auditory stimuli and bi-manual responses, the responsive
regions included bilateral visual occipito-temporal cortices,
bilateral auditory cortices, motor, premotor and cerebellar cortices,
and a large-scale bilateral parieto-frontal structure.  In the present
analysis we follow \cite{Gallos2012} and we do not explore the
differences in networks between different SOA conditions.  Rather, we
consider them as independent equivalent experiments, generating a
total of 64 different scans, one for each condition of temporal gap
and subject.

The following emergent clusters are seen in resting state: medial
prefrontal cortex, posterior cingulate, and lateral temporoparietal
regions, all of them part of the default mode network (DMN) typically
seen in resting state data and specifically found in our NYU dataset
\cite{Shehzad2009}.


\subsection{Computation of parameters $\gamma$, $\alpha$, $\beta$, and 
$k_{\rm max}$}
\label{ks}

Once $T_c$ is determined, we are able to compute the degree
distribution of the brain networks.  For a given brain scan we search
for all connected components of strong links with $C_{ij}>T_c$, where
$T_c$ is the first jump in the largest connected component as seen in
Fig. \ref{fig3.fig}. We then calculate $P(k_{\rm in})$ using all brain
networks for a given experiment; the results are plotted in
Fig. \ref{fig3.fig}.  We consider all nodes with $k_{\rm in}\geq 1$ at
$T_c$ from all the connected clusters.  As one can see in
Fig.~\ref{fig3.fig}b, for all data sets, we found degree distributions
which can be described by power laws $P(k_{\rm in})\sim k_{\rm
in}^{-\gamma}$ with a given cut-off $k_{\rm max}$.  For the resting
state , we found $\gamma=2.85\pm 0.04$ and $k_{\rm max}=133$ while for
the dual task we found $\gamma=2.25\pm 0.07$, $k_{\rm max}=139$
(see Table \ref{exp.table}).  We use a statistical test based on
maximum likelihood methods and bootstrap analysis to determine the
distribution of degree of the networks.  We follow the method of
Clauset, Shalizi, Newman, SIAM Review {\bf 51}, 661 (2009) of maximum
likelihood estimator for discrete variables which was already used in
our previous analysis of the dual task data
\cite{Gallos2012}.


We fit the degree-distribution assuming a power law within a given
interval. For this, we use a generalized power-law form
\begin{equation}
  P(k;k_{\rm min},k_{\rm max})
  = \frac{k^{-\gamma}}{\zeta(\gamma,k_{\rm min})-\zeta(\gamma,k_{\rm max})} ,
\end{equation}
where $k_{\rm min}$ and $k_{\rm max}$ are the boundaries of the fitting
interval and the Hurwitz $\zeta$ function is given by
$\zeta(\gamma,\alpha)= \sum_i (i+\alpha)^{-\gamma}$. We set
$k_{\rm min}=1$.

We calculate the slopes in successive intervals by continuously
increasing $k_{\rm max}$. For each one of them we calculate the
maximum likelihood estimator through the numerical solution of
\begin{equation}
  \gamma = {\rm argmax} \left( -\gamma \sum_{i=1}^M \ln k_i -
  M \ln \left[ \zeta(\gamma,k_{\rm min})-\zeta(\gamma,k_{\rm
  max}) \right] \right),
\end{equation}
where $k_i$ are all the degrees that fall within the fitting interval
and $M$ is the total number of nodes with degrees in this interval.
The optimum interval was determined through the Kolmogorov-Smirnov
test.

For the goodness-of-fit test, we use 
KS test 
generating 10,000 synthetic random distributions following the
best-fit power law. 
Analogous analysis is performed to test for a possible exponential
distribution to describe the data.
We use KS statistics to determine the optimum fitting intervals and
also the goodness-of-fit.  In all the cases where the power law was
accepted
we ruled out the possibility of an exponential distribution,
see \cite{Gallos2012}.



In order to compute the correlation of $k_{\rm in}$, $k_{\rm out}$ and
$k_{\rm in}^{\rm nn}$ we consider the following statistics for the
weak links and the degrees of the external nearest neighbors of an
outgoing node.
This correlation is gathered from the calculation of the average
in-degree, $\av{k^{\rm nn}_{\rm in}}$ of the external neighbors of a
node with in-degree $k_{\rm in}$. The strong-links are those links
added to the network for $T > T_c$. The weak links are those added to
the network for values of $T_0 < T < T_c$ until the average out-degree
reaches $\langle k_{\rm out}\rangle = 1$.  For statistical
determination of the scaling properties of weak-links, we consider
that they connect two nodes in different networks, or even nodes in
the same component.  To calculate the statistical scaling properties
of weak links, we consider the out-weak-degree $k_{\rm out}$ of a node
as the number of all links added for $T_0<T<T_c$.

Figure~\ref{fig3.fig}f shows that the scenario for the correlation
between $\av{k^{\rm nn}_{\rm in}}$ and $k_{\rm in}$
is consistent with Eq. ~(\ref{beta.eq}).
For the resting state experiments (Fig.~\ref{fig3.fig}f) there is a
positive correlation between the $k_{\rm in}$ of outgoing nodes placed
in different functional networks. For the dual-task human subjects
(Fig.~\ref{fig3.fig}f) the correlation is also positive.

Moreover, when analyzing the relation between $k_{\rm in}$ and $k_{\rm
out}$ for the same outgoing nodes, they are described by the
correlations presented in Fig.~\ref{fig3.fig}e using power laws.
Figures ~\ref{fig3.fig}e-f depict the power-law fits using Ordinary
Least Square method within a given interval of degree.  We assess the
goodness of fitting in each interval via the coefficient of
determination $R^2$. We accept fittings where $R^2 \gtrsim 0.9$. The
exponents measured are presented in Table~\ref{exp.table}.

Figures~\ref{fig4.fig}a and b show the results we found when we apply
the theory presented in Section \ref{generating} of this Supplementary
Information on two coupled networks of degree exponent $\gamma=2.85$ and
$2.25$, respectively with the cut-off given by $k_{\rm max}=133, 139$,
respectively as given by the values for human resting state and dual
task.  For $\gamma=2.25$ and $\gamma=2.85$, the value associated
with the data gathered from humans, the results are similar with those
presented on Fig.~\ref{fig2.fig} in both theoretical cases, the
conditional (left panels ) and redundant (right panels) interactions.
The main differences between the results for $\gamma=2.25$ and $2.85$
are the values found for $p_c$, where the values found for
$\gamma=2.25$ are systematically smaller than the values found for
$\gamma=2.85$, going from $p_c\approx 0.1$ to $\approx 0.6$ for
$\gamma=2.25$, and from $p_c\approx 0.1$ to $\approx 0.8$ for
$\gamma=2.85$. These results can be understood from the knowledge
gathered on the percolation of single networks~\cite{Cohen2002}. For
lower values of the degree exponent $\gamma$ the hubs on scale-free
networks become more frequent, protecting the network from breaking
apart.
When comparing the two cases of Fig.~\ref{fig4.fig} with the
theoretical case of $\gamma=2.5$ (Fig.~\ref{fig2.fig}), one can notice
that the broader the distribution (as lower the value of $\gamma$),
the more robust is the system of coupled networks.  There general
trends are consistent with the calculations of $p_c$ for unstructured
interconnected networks with one-to-one connections done in Ref.
\cite{Buldyrev2010}. 
The white circles in Fig.~\ref{fig4.fig} correspond to the values of
$\alpha$ and $\beta$ measured from real data. As one can see, the
experimental values are placed on the region that represents the best
compromise between the predictions for optimal stability under
conditional and redundant interactions.

It is also interesting to note that the extreme vulnerability
predicted in Ref. \cite{Buldyrev2010} can be somehow mitigated by
decreasing the number of one-to-one interconnections as shown in
Parshani, R., Buldyrev, S. V. \& Havlin, S.  Interdependent networks:
reducing the coupling strength leads to a change from a first to
second order percolation transition.  {\it Phys. Rev. Lett.} {\bf
105}, 048701 (2010).
However, in this case, the system of networks may be rendered
non-operational due to the lack of interconnections. Indeed, by
connecting both networks with one-to-one outgoing links and by making
these interconnections at random, there is a high probability that a
hub in one network will be connected with a low degree node in the
other network. These low degree nodes are highly probable to be chosen
in a random attack, thus the hubs become very vulnerable due to the
conditional interaction with a low degree node in the other
network. This effect leads to the catastrophic cascading behavior
found in \cite{Buldyrev2010}.

Another way to protect a network in the conditional mode is to
increase the number of out-going links per nodes, since the failure of
a node occurs when all its inter-linked nodes have failed. Thus, by
just increasing the number of interlinks from one to many out-going
links emanating from a given node, larger resilience is obtained. If
these links are distributed at random, then this situation corresponds
to $\alpha=\beta=0$ in our model.  However, in this random conditional
case, the network may be rendered non-operational due to the random
nature of the interlink connectivity. A functional real network is
expected to be operating with correlations and therefore the most
efficient structure when there are many correlated links connecting
the networks is the one found for the brain networks investigated in
the present work.  In other words, assuming that a natural system like
the brain functions with intrinsic correlations in inter-network
connectivity, then the solution found here (large $\alpha$ and
$\beta>0$) seems to be the natural optimal structure for global
stability and avoidance of systemic catastrophic cascading effects.

Another problem of interest is the targeted attack of interdependent
networks as treated in Huang, X., Gao, J., Buldyrev, S. V., Havlin,
S. \& Stanley, H. E.  Robustness of interdependent networks under
targeted attack.  {\it Phys. Rev. E} {\bf 83}, 065101 (2011).
It would be of interest to determine how the present correlations
affect the targeted attack to, for instance, the highly connected
nodes.


\newpage

%

%

%
%


\end{document}